\documentclass[prd,twocolumn,nofootinbib,showpacs]{revtex4-1}

\usepackage{amsfonts}
\usepackage{yfonts}

\expandafter\let\csname equation*\endcsname\relax
\expandafter\let\csname endequation*\endcsname\relax
\usepackage{amsmath}

\usepackage{amssymb}
\usepackage{bm}
\usepackage{dcolumn}
\usepackage{graphicx}
\usepackage{graphics}
\usepackage[latin1]{inputenc}
\usepackage{latexsym}
\usepackage{rotating}
\usepackage{hyperref}
\usepackage[all]{hypcap} 
\usepackage{xspace} 
\usepackage[usenames,dvipsnames]{color}
\usepackage{mathrsfs}
\usepackage{subfigure}
\usepackage{soul}

\usepackage{ulem}
\usepackage{outlines}
\usepackage{enumitem}
\setenumerate[1]{label=\Roman*.}
\setenumerate[2]{label=\Alph*.}
\setenumerate[3]{label=\roman*.}
\setenumerate[4]{label=\alph*.}

\definecolor {darkgreen}{rgb}{0.2,0.7,0.2}

\usepackage{etoolbox}

\makeatletter
\def\@mkboth#1#2{}
\newlength\appendixwidth
\preto\appendix{\addtocontents{toc}{\protect\patchl@section}}
\newcommand{\patchl@section}{%
  \settowidth{\appendixwidth}{\textbf{Appendix }}%
  \addtolength{\appendixwidth}{1.5em}%
  \patchcmd{\l@section}{1.5em}{\appendixwidth}{}{\ddt}%
}
\makeatother

\newcommand\be{\begin{equation}}
\newcommand\ba{\begin{eqnarray}}
\newcommand\ee{\end{equation}}
\newcommand\ea{\end{eqnarray}}

\newcommand\bw{\begin{widetext}}
\newcommand\ew{\end{widetext}}

\newcommand{\nn}{\nonumber}

\newcommand*\pFqskip{8mu}
\catcode`,\active
\newcommand*\pFq{\begingroup
        \catcode`\,\active
        \def ,{\mskip\pFqskip\relax}%
        \dopFq
}
\catcode`\,12
\def\dopFq#1#2#3#4#5{%
        {}_{#1}F_{#2}\biggl[\genfrac..{0pt}{}{#3}{#4};#5\biggr]%
        \endgroup
}

\begin{document}
\title{Fast Frequency-Domain Effective Fly-By Waveforms}

\author{Nicholas Loutrel}
\address{Department of Physics, Princeton University, Princeton, NJ, 08544, USA}

\date{\today}

\begin{abstract} 
Recently, we developed effective fly-by (EFB) waveforms designed to model the burst of gravitational radiation from highly eccentric binaries. We here present a faster to evaluate frequency domain EFB waveform. The waveform is constructed through the use of asymptotic expansions of hypergeometric functions. Since the waveform is fully analytic, we study the accuracy to which the binary's parameters can be measured using a Fisher analysis. We find that degeneracies exist among the parameters, such that the waveform is parameterized in terms of the chirp mass ${\cal{M}}$ and orbital radius of curvature ${\cal{P}}$, instead of the total mass, symmetric mass ratio, and semi-latus rectum of the binary. By computing the Fisher matrix for single bursts from two thousand binary systems, we find that most of the systems will have greater than one hundred percent uncertainty in the chirp mass, luminosity distance, and inclination angle, while roughly half will have less than one hundred percent uncertainty in the orbital radius of curvature, orbital eccentricity, and polarization angle. Further, we repeat this analysis after including additional bursts within the inspiral sequence and find that the uncertainties in the waveform's parameters can improve by orders of magnitude with a sufficient timing model for the bursts.
\end{abstract}



\maketitle


\newpage
\section{Introduction}

Over the past few decades, an industry has been built surrounding the modeling of gravitational waves (GWs) from compact binary systems. Many methods are used to develop model of GWs, namely (but not limited to) numerically solving the full Einstein field equations~\cite{Bishop:2016lgv}, perturbatively solving the equations at small velocities~\cite{Blanchet:2013haa} or mass ratios~\cite{Barack:2018yvs}, using perturbation theory to solve for the linearized response of black holes~\cite{Teukolsky:1973ap}, and treating the problem through effective spacetimes~\cite{Damour:2011xga}. The efficacy of these methods have been tested with each detection by the Laser Interferometer Gravitational wave Interferometer (LIGO)~\cite{Abramovici:1992ah, Harry:2010zz, TheLIGOScientific:2014jea} and Virgo~\cite{Caron:1997hu, TheVirgo:2014hva} observatories. With more detectors planned and proposed, these methods, as well as new ones, will continue to be useful tools for studying compact binary coalescences.

A subset of binary systems that has historically been neglected within this industry, but has received renewed interest in recent years, are eccentric binaries. Binaries can enter the detection band of ground based detectors with non-negligible eccentricities though a few mechanism, namely dynamical interactions in dense stellar environments~\cite{Samsing:2013kua, Rodriguez:2018pss, Samsing:2017xmd, Samsing:2018ykz, Samsing:2019dtb, 2012PhRvD..85l3005K, 2009MNRAS.395.2127O, Leigh:2017wff, Miller:2009wv} and hierarchical triple systems~\cite{Antognini:2013lpa, Antonini:2015zsa}. A small subset of these systems will enter the detection band with large eccentricity, close to the unbound limit. The gravitational waves from such systems resemble a sequence of bursts, rather than the continuous chirping signal of quasi-circular systems.

Despite being historically overlooked due to an incomplete picture of formation channels and questions regarding detectability, eccentric binaries may be a useful tool to understanding fundamental physics with GW detections. Since the pericenter velocity of these systems can be large, they present themselves as an interesting laboratory for tests of general relativity. Eccentricity has been shown to have a non-trivial impact on constraints of modified theories of gravity~\cite{Moore:2020rva, Ma:2019rei, Loutrel:2014vja}. Further, if one (or both) of the binary components is a neutron star, f-modes can be excited on the star due to tidal forces during closest approach. These effects may prove to be a powerful tool for constraining the equation of state of dense nuclear matter~\cite{Yang:2018bzx, Yang:2019kmf, 2019arXiv191204892V, Vick:2019cun, 2018MNRAS.476..482V}.

\begin{figure}
	\includegraphics[scale=0.3]{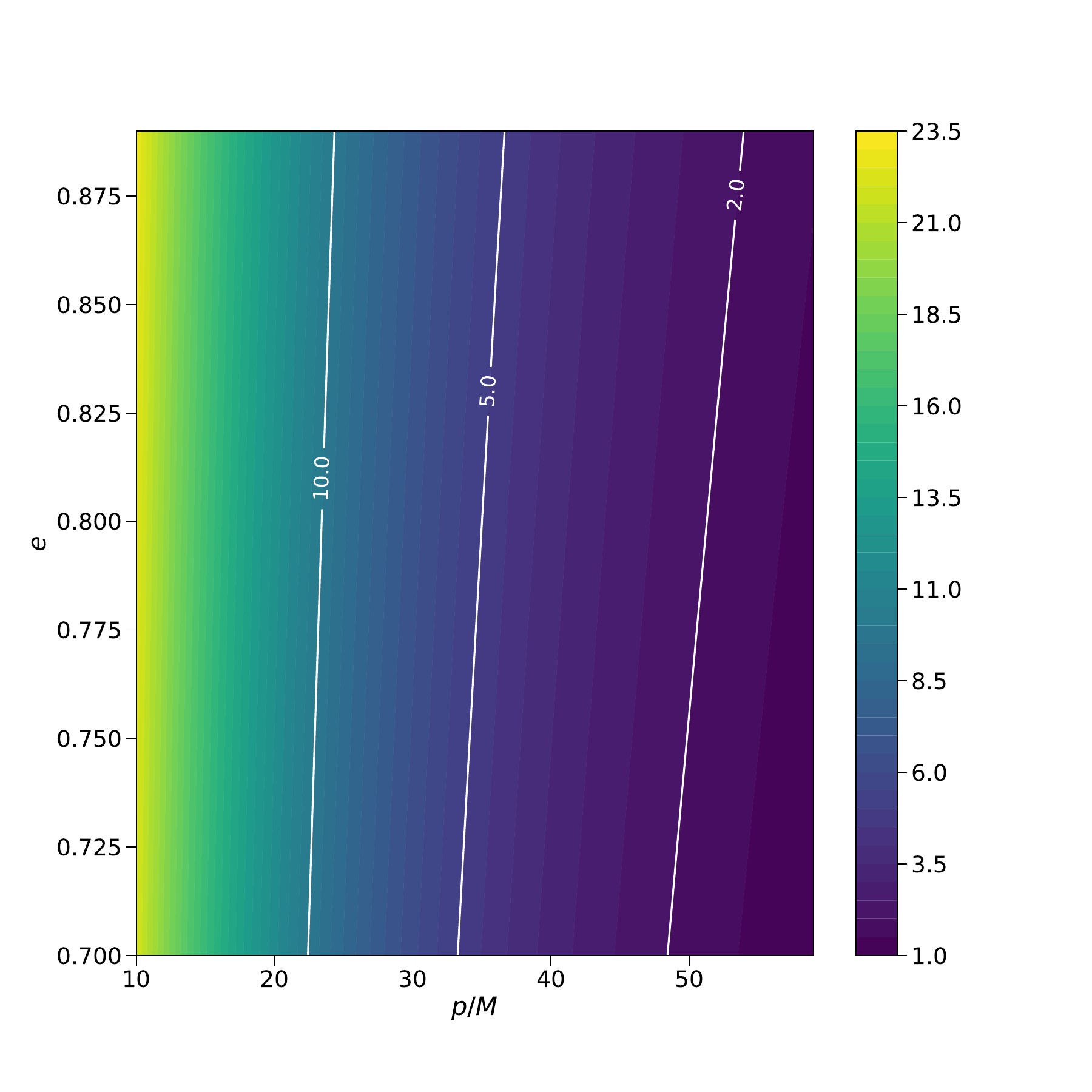}
	\caption{\label{snr} Signal to noise ratio (SNR), as calculated via the method described in the text, of a gravitational wave burst from a single pericenter passage of binary systems with varying values of the orbital eccentricity $e$ and semi-latus rectum $p$. The remaining parameters are fixed at masses of $(10, 10) M_{\odot}$, luminosity distance of $100$ Mpc, and the system is oriented in the ``face-on" configuration with polarization angle $\beta = 0$. The waveforms are computed by numerically integrating the equations governing Newtonian orbits combined with the equations for the leading PN order radiation reaction effects, an approximation commonly called the Newtonian plus quadrupole radiation approximation. The SNR shows little variation as the eccentricity increase, but can vary by more than an order of magnitude as the semi-latus rectum is decreased, making the binary more compact.}
\end{figure}

A point that is commonly made about the burst phase of eccentric binary systems is that the bursts themselves are weak, having low signal-to-noise ratio (SNR), and will be difficult to detect. However, this statement is strongly dependent on the separation of the binary at closest approach, or alternatively, on the semi-latus rectum $p$ of the orbit. The SNR of a single burst generated during a single pericenter passage can vary by more than an order of magnitude, as can be seen from Fig.~\ref{snr}. Although rare, it is thus not implausible to expect GW bursts from highly eccentric systems with reasonable SNRs. This opens the door to searching for such signals using matched filtering techniques, provided one can create a sufficiently accurate waveform for the bursts.

Within the past several years, there have been many advances toward developing models of the gravitational wave emission from eccentric binaries. A few of these are the third post-Newtonian (3PN) inspiral waveform of~\cite{Moore:2019xkm} valid for eccentricities $e\lesssim0.8$, the hybrid inspiral-merger-ringdown ENIGMA model~\cite{Huerta:2017kez} valid for $e\lesssim0.4$, and the SEOBNRE model~\cite{Cao:2017ndf} within the effective one-body framework valid for $e\lesssim0.6$. More recently, a numerical model for dynamical capture binaries was developed within the effective-one-body formalism in~\cite{Nagar:2020xsk}, constituting one of the few models for the high eccentricity regime $(e\sim1)$.

Recently, we developed the first analytic waveforms for the GW bursts of highly eccentric binaries~\cite{Loutrel:2019kky}. The waveforms were computed by working in the Newtonian plus quadrupole radiation paradigm, where the conservative dynamics of the binary are modeled using Newtonian orbits, while radiation reaction was computed using the leading PN order quadrupole approximation. The models were developed by performing a re-summation of common Fourier series representations of the orbital dynamics, resulting in an effective fly-by (EFB) approach. Focusing on the time domain, the EFB-T waveform was shown to be an accurate representation of leading PN order burst waveforms, while still retaining reasonable agreement with numerical relativity fly-by waveforms.

However, a problem arose when moving to the Fourier domain. The frequency domain waveform, called the EFB-F model, took several hours to evaluate for a single system, rendering it completely impractical for any real searches. We here present a new frequency domain EFB waveform, called the EFB-F2 model, which takes less time to evaluate than the original EFB-F model by about 4-5 orders of magnitude. The model is achieved through a rather lengthy procedure, but there are a few critical steps. The first is a linear transformation of the hypergeometric functions that the original EFB-F model depends on. This transformation effectively allows us to perform a post-Newtonian (PN) expansion of these functions. Second, is an asymptotic expansion of the PN-expanded hypergeometric functions at large frequencies. Afterwards, we re-sum the PN-expansion of the now doubly expanded hypergeometric functions in terms of Bessel functions of the first kind. By computing the match between the EFB-F2 model and numerical leading-PN order burst waveforms, we find that the new model is a faithful representation of eccentric burst signals for a wide region of parameter space (see Fig.~\ref{match}).

While the new waveforms may be faster, the fact that they are also analytic opens the door to performing a parameter estimation study using a Fisher analysis~\cite{Finn:1992xs, cutlerflanagan}. This method relies on assuming that the expected signal is sufficiently similar to the waveform model that one can approximate the log-likelihood as a quadratic function in the parameters of the model, with coefficients that depend on the derivatives of the model with respect to the parameters. The uncertainties in waveform paramters are then found by inverting the matrix of coefficients. We carry out this analysis with the EFB-F2 waveforms for single burst events, as well as multi-burst inspiral sequences. For the single burst analysis, we fix the masses to be $10 M_{\odot}$ and luminosity distance to the source to be $10$ Mpc. We then generate two thousand systems by randomly sampling the semi-latus rectum, orbital eccentricity, inclination angle, polarization angle, and sky location of the source. 

We have found that the EFB-F2 waveform model actually has degeneracies among its parameters, such that the waveform is parameterized in terms of the chirp mass ${\cal{M}} = M \eta^{3/5}$ and orbital radius of curvature ${\cal{P}} = (p^{3}/M)^{1/2}$, with $M$ the total mass of the binary, $\eta$ the symmetric mass ratio, and $p$ the semi-latus rectum of the orbit. After computing the Fisher matrix for each of these systems, we find that the uncertainties for many of the systems are typically larger than one hundred percent. For example, nearly all of the systems studied have more than one hundred percent uncertainty in the chirp mass and luminosity distance. These two parameters also show a large correlation. Meanwhile, approximately half of the systems studied have less than one hundred percent uncertainty in the orbital radius of curvature and orbital eccentricity, which also show very strong correlations.

We repeat this analysis for all of the two thousand systems after including the burst following the initial in the inspiral sequence. The Fisher analysis predicts that the uncertainties in parameters can improve by an order of magnitude or more simply by including an additional burst. We study the behavior of this trend with the number of pericenter passages, and thus bursts, by selecting one system and repeating the Fisher analysis for a total of twenty five passages. While the Fisher analysis predicts orders of magnitude improvements in the uncertainties, after a few bursts, the uncertainties converge to a monotonically decreasing trend with only small improvements from one burst to the next.

 This paper is organized as follows. In Sec.~\ref{ffw}, we explain the steps necessary to computed the EFB-F2 model from the original EFB-F model, with a derivation of the asymptotic expansion of the hypergeometric functions given in Appendix~\ref{2f1-asym}. The Fisher analysis for single bursts is carried out in Sec.~\ref{single}, while the multi-burst analysis is discussed in Sec.~\ref{multi}, with the main results presented in Figs.~\ref{fisher-plot}-\ref{fisher-plot2}. Finally, in Sec.~\ref{discuss}, we discuss some of the possible pitfalls of the Fisher analysis carried out herein. Throughout this work, we use $G = c = 1$.
 
\section{Fast Fourier Waveforms}
\label{ffw}

We here present the necessary details needing to construct the EFB-F2 model, and its accuracy compared to numerical waveforms.

\subsection{Review of the EFB-F Waveform}

Before we discuss the procedure by which we obtain the EFB-F2 waveforms, it is useful to review the EFB-F waveform from which they are derived~\cite{Loutrel:2019kky}. The starting point for these waveforms was writing the time domain polarizations $h_{+,\times}(t)$ as Fourier series of the orbital frequency. A radiation reaction model was developed that described the changes in orbital elements, specifically the semi-latus rectum $p$ and eccentricity $e$, as well as mean anomaly $\ell$ over one pericenter passage. Combining the radiation reaction model with the Fourier series representation, we were able to compute the Fourier transform of the waveform polarizations using the stationary phase approximation~\cite{Bender}, obtaining the frequency domain polarizations $\tilde{h}_{+,\times}(f)$. These frequency domain polarizations were still written as infinite summations on specialized Bessel functions, known specifically as Kapteyn series. The series themselves were re-summed by replacing the Bessel functions with their asymptotic representations, and taking the summation to an integral, which could be analytically evaluated in closed-form. The waveforms that resulted from this procedure take the form
%
\begin{align}
\tilde{h}_{+,\times}^{\rm EFB-F} = h_{0} {\cal{A}}_{0}(f) \sum_{(l_{1}, l_{2}) \in L} \sum_{s} h_{l_{1}, l_{2}, s}(f)
\end{align}
%
where 
\begin{align}
h_{0} &= \frac{M^{2} \eta}{p_{0} D_{L}} \frac{(1-e_{0}^{2})}{e_{0} F_{\rm rr}}
\\
{\cal{A}}_{0}(f) &= \left(\frac{\chi}{\chi_{\rm orb}}\right)^{i\chi} \frac{e^{2\pi i f t_{p} - i \chi}}{\chi^{1/2}}
\\
\label{eq:hls}
h_{l_{1}, l_{2}, s} &= {\cal{A}}_{l_{1}, l_{2}, s}(f) \; \pFq{2}{1}{\frac{l_{1}}{6}-i \frac{\chi}{2},\frac{l_{2}}{6} - i \frac{\chi}{2}}{s}{X}
\end{align}
with $M$ the total mass of the binary, $\eta$ the symmetric mass ratio, $(p_{0}, e_{0})$ the semi-latus rectum and eccentricity at pericenter, $D_{L}$ the luminosity distance, $t_{p}$ the time of pericenter passage, $\chi = f/F_{\rm rr}$, $\chi_{\rm orb} = n_{0}/2\pi F_{\rm rr}$, and $X = -(9/4) \chi_{\rm orb}^{2}/\zeta_{0}^{3}$. The parameters $(n_{0}, F_{\rm rr}, \zeta_{0})$ are known functions of the binary's masses and orbital parameters, specifically
\allowdisplaybreaks[4]
\begin{align}
n_{0} &= M^{1/2} \left(\frac{1-e_{0}^{2}}{p_{0}}\right)^{3/2}\,,
\\
F_{\rm rr} &= \frac{96}{10\pi} \frac{\eta}{M} \left(\frac{M}{p_{0}}\right)^{4} \left(1 - e_{0}^{2}\right)^{1/2} \left(1 + \frac{73}{24} e_{0}^{2} + \frac{37}{96} e_{0}^{4}\right)\,,
\\
\zeta_{0} &= \left\{\frac{3}{2} \left[\ln\left(\frac{1 + \sqrt{1-e_{0}^{2}}}{e_{0}}\right) - \sqrt{1-e_{0}^{2}}\right]\right\}^{2/3}\,.
\end{align}
The summations indices $(l_{1}, l_{2})$ belong to the set $L = \{(2,4), (4,8), (1,5), (5,7), (7,11), (10,8)\}$, while $s \in \{-1/2,1/2\}$. The function $_{2}F_{1}$ is the Gauss hypergeometric functions, while the amplitude functions ${\cal{A}}_{l_{1}, l_{2}, s}$ are given in Appendix C of~\cite{Loutrel:2019kky}.

While it is instructive to be able to make analytic Fourier domain waveforms, the EFB-F model is impractical due to its excessively long evaluation time. In \texttt{Python}, we estimate it will take $\sim$10 hours to generate a single EFB-F waveform using the methods described in~\cite{Loutrel:2019kky}. The reason for this is two fold. First, evaluating specialized functions numerically is generally slow, especially in regions of parameter space where typical methods may be slowly convergent. Second, is the need for arbitrary floating point precision. The hypergeometric functions in Eq.~\eqref{eq:hls} actually grow exponentially with $\chi$ and become sufficiently large that they cannot be evaluated at double precision. Of course, if these issues can be alleviated, then the waveform becomes viable for use is searches and parameter estimation. We here provide an analytic treatment that significantly speeds up the evaluation of the waveform. We shall call these new waveforms the EFB-F2 model.

\subsection{EFB-F2 Waveform}

To obtain the new, faster to evaluate waveform, we use the following procedure:
\begin{enumerate}
	\item Apply the hypergeometric function transformation $X \rightarrow 1/(1-X)$, specifically~\cite{NIST}
\begin{align}
\label{eq:2f1-id}
\pFq{2}{1}{a,b}{c}{X} &= \frac{\gamma(a,b,c)}{(1-X)^{a}} \pFq{2}{1}{a, c-b}{a-b+1}{\frac{1}{1-X}} + \left(a \leftrightarrow b\right)\,,
\\
\gamma(a,b,c) &= \frac{\Gamma(c) \Gamma(b-a)}{\Gamma(b) \Gamma(c-a)}\,.
\end{align}

	\item Replace the hypergeometric functions of argument $1/(1-X)$ with their asymptotic expansion, derived explicitly in Appendix~\ref{2f1-asym}.
	\item Replace the $\gamma$ function with its asymptotic expansion about $\chi\gg1$ and define $\tilde{\gamma}_{l_{1}, l_{2}, s}$ such that
\begin{align}
\gamma\left(\frac{l_{1}}{6} - i \frac{\chi}{2}, \frac{l_{2}}{6} - i \frac{\chi}{2}, s\right) \sim e^{\pi \chi/2} \tilde{\gamma}_{l_{1}, l_{2}, s}(\chi)\,.
\end{align}

	\item Replace the amplitude functions ${\cal{A}}_{l_{1}, l_{2}, s}$ with their asymptotic expansions about $\chi\gg1$ and define $\tilde{{\cal{A}}}_{l_{1}, l_{2}, s}$ such that
\begin{align}
{\cal{A}}_{l_{1}, l_{2}, s}(f) \sim e^{-\pi \chi/2} \tilde{{\cal{A}}}_{l_{1}, l_{2}, s}(f)\,.
\end{align}

	\item Cancel the exponential growth in $\gamma$ with the exponential decay in ${\cal{A}}$.
	\item Combine the previous steps together to create the waveform polarizations, and apply a high frequency cutoff.
\end{enumerate}
This procedure results in the waveform
\begin{widetext}
\begin{equation}
\label{eq:efb-f2}
\tilde{h}_{+,\times}^{\rm EFB-F2} = h_{0} {\cal{A}}_{0}(f) \Theta(f_{\rm cut} - f) \sum_{(l_{1}, l_{2}) \in L} \sum_{s,n} \tilde{{\cal{A}}}_{l_{1}, l_{2}, s}(f) \left[\frac{\tilde{\gamma}_{l_{1},l_{2},s}(f)}{(1-X)^{\frac{l_{1}}{6} - i \frac{\chi}{2}}} {\cal{J}}_{n}(l_{1}, l_{2}, s; \chi) J_{\frac{1}{6}(l_{1}-l_{2})+n}\left(\frac{2 e^{{\cal{G}}_{1}/2}}{\sqrt{X-1}}\right) + (l_{1} \leftrightarrow l_{2})\right]
\end{equation}
\end{widetext}
where the ${\cal{J}}_{n}$ and ${\cal{G}}_{1}$ functions are given in Appendix~\ref{2f1-asym}, and the $\tilde{{\cal{A}}}_{l_{1},l_{2},s}$ are given in Appendix~\ref{amps}. Before we provide explicit details of these steps, we will provide the reasoning behind them.

Our starting point is the realization that $\chi \gg 1$ since $F_{\rm rr} \ll 10$ Hz for most sources of ground based-detectors, and that $X$ scales like $v^{-5}$ and is in the range $X \in (-\infty, 0]$. The former of these implies that we can work in an asymptotic expansion about $\chi$ being large, while the latter allows us to employ the identity in Eq.~\eqref{eq:2f1-id}. While investigating this, we discovered that the identity is only exact numerically when $(a,b)$ are real valued. When they are complex, as they are in our application, the identity is only approximate, diverging from the original hypergeometric function on the left hand side of Eq.~\eqref{eq:2f1-id}. This issue is purely numerical, and is not an analytic property of the hypergeometric functions. The divergence appears to arise due to both the left and right hand sides of Eq.~\eqref{eq:2f1-id} growing exponentially large, to the point that they are not well estimated at double precision accuracy. By evaluating at higher precision using \texttt{mpmath} in \texttt{Python}, the error between the left and right hand sides of Eq.~\eqref{eq:2f1-id} decreases. However, going to higher precision slows down the evaluation of the hypergeometric functions, so there is a trade-off between speed and accuracy. From a practical standpoint associated with this model, this numerical error produces an exponential growth in the EFB-F2 waveform at sufficiently high frequencies, which is not present in the exact answer (i.e. those generated via numerical evolution of the PN equations of motion). To correct for this, we apply a high frequency cutoff to the model, which should not result in a significant loss of power since the bursts exponentially decay in frequency.

The exact value of the frequency cutoff depends on the parameters of the binary. Lower values of the semi-latus rectum, and higher values of the eccentricity, require a cutoff at higher frequencies. To determine a suitable cutoff frequency, we generated thirty-five EFB-F2 waveforms, located the turning point between exponential decay and exponential growth ``by eye," and selected the cutoff frequency to be 10 Hz below this. From this data, we generated a fitting functions of the form
\begin{equation}
\label{eq:fcut}
f_{\rm cut} = \frac{M^{-1}}{\bar{p}^{F} (1-e_{0}^{2})^{E}} \left(A + B e_{0} + C e_{0}^{2} + D e_{0}^{3}\right)\,,
\end{equation}
where $\bar{p} = p/M$, and with the coefficients
\begin{align}
A &= 2.24674793\,, \qquad B = -7.44133144\,,
\\
C &= 13.10225776\,, \qquad D = -7.90769141\,,
\\
E &= 1.08108155\,, \qquad F = 1.27977090\,.
\end{align}

The second step involves asymptotically expanding the hypergeometric functions of argument $1/(1-X)$ on the right hand side of Eq.~\eqref{eq:2f1-id} about $\chi \gg 1$. Since $X \sim v^{-5}$, the argument $1/(1-X)$ is small and one can employ the well known hypergeometric series
\begin{equation}
\label{eq:2f1-series}
\pFq{2}{1}{a, c-b}{a-b+1}{\frac{1}{1-z}} = \sum_{j=0}^{\infty} \frac{(a)_{j} (c-b)_{j}}{j! (a-b+1)_{j}} \left(\frac{1}{1-z}\right)^{j}\,.
\end{equation}
where $(a)_{j}$ is the Pochhammer symbol. The series expansion effectively constitutes a PN expansion. Since we are working to leading PN order, then it might seem logical to simply truncate the sum at leading order in $v$. However, we found that this results in a significant loss of accuracy compared to an exact answer. To retain accuracy, we expand this series about $\chi \gg 1$, and re-sum it. The exact process of this will be detailed in Appendix~\ref{2f1-asym}. Eliminating these hypergeometric functions from the model in this fashion results in a 3-4 order of magnitude speed up when evaluating the model, and does not result in a significant loss of accuracy.

The next two steps involve the asymptotic expansions of the functions $\gamma$ and ${\cal{A}}$. These functions depend on the Gamma function, whose asymptotic expansion is well known. From this expansion we can factor out $e^{\pi \chi/2}$ from $\gamma$ and $e^{-\pi \chi/2}$ from ${\cal{A}}$. This step is crucial for making these waveforms fast to evaluate. Since $\chi \gg 1$, these exponential factors can be sufficiently large or small that one needs more than double precision accuracy to evaluate them, resulting in a slow down when evaluating the model. Properly factoring these out and cancelling them results in a 1-2 order of magnitude speed up in evaluating the model.

The exact evaluation time of the EFB-F2 waveform polarizations given in Eq.~\eqref{eq:efb-f2} depends on the desired frequency resolution. In \texttt{Python}, at a resolution of $\delta f = 0.1$ Hz, and sampling from $f_{\rm low} = 10$ Hz to $f_{\rm high} = 2048$ Hz, it takes roughly $0.26$ seconds to evaluate the plus polarization with parameters $p_{0} = 20 M$, $e_{0} = 0.9$, and $\iota = 0 = \beta$. For a higher resolution of $\delta f = 0.01$ Hz, which is roughly what was used in~\cite{Loutrel:2019kky}, it takes ten times longer to evaluate the model. This is still slow compared to the EFB-T model developed in~\cite{Loutrel:2019kky}, but it is 4-5 orders of magnitude faster than the original EFB-F model.

\begin{figure}
	\includegraphics[scale=0.3]{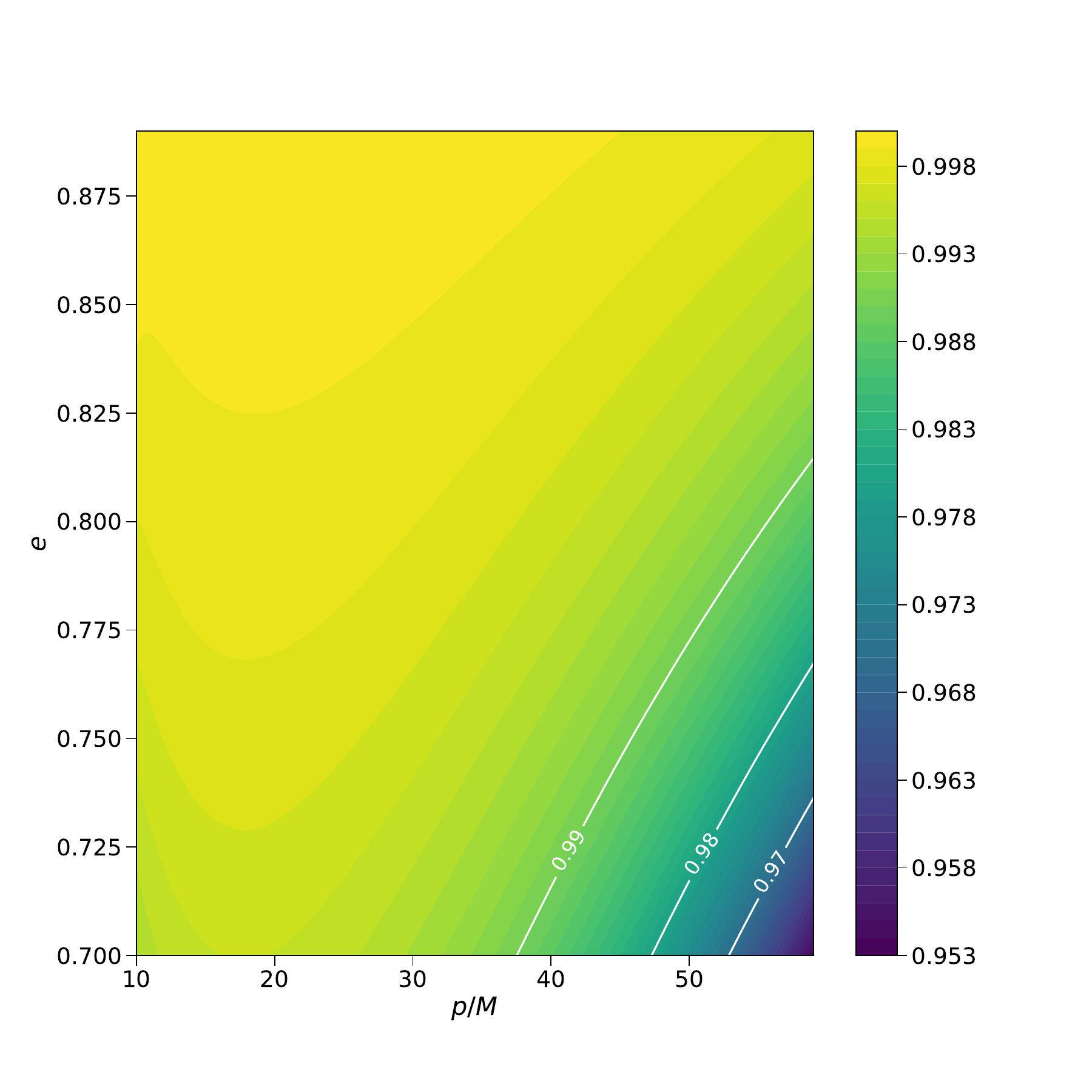}
	\caption{\label{match} Match (color) between the EFB-F2 waveforms and numerical leading PN order burst waveforms as a function of the semi-latus rectum $p$ and eccentricity $e$. The white lines show the contours corresponding to matches of 0.97, 0.98, and 0.99, respectively.}
\end{figure}

To study how faithful an approximation the EFB-F2 waveforms are to an exact answer, in this case a numerical leading PN order burst waveform, we compute the match given by~\cite{Buonanno:2009zt}
\begin{equation}
\text{M} = \underset{t_{p}}{\text{max}} \frac{\left(h_{\rm num} | h_{\rm EFB}\right)(t_{p})}{\left(h_{\rm num} | h_{\rm num}\right)^{1/2} \left(h_{\rm EFB} | h_{\rm EFB}\right)^{1/2}}
\end{equation}
where $\left(A|B\right)$ is the noise weighted inner product between waveforms $A$ and $B$
\begin{equation}
\left(A | B\right) = 4 \text{Re} \int_{0}^{\infty} df \frac{\tilde{A}(f) \tilde{B}^{\dagger}(f)}{S_{n}(f)}\,,
\end{equation}
with $S_{n}(f)$ the noise spectral density of the detector and $\dagger$ corresponding to complex conjugation. For $S_{n}(f)$, we use the design sensitivity curve provided in~\cite{LIGOsn}. The match is in the range $[0,1]$, and allows us to quantify the error our approximations have introduced relative to a detector's sensitivity. The numerical leading PN order waveforms $h_{\rm num}$ are generated by numerical integration of the leading order PN equations of motion, which is described in more detail in~\cite{Loutrel:2019kky}. The results of the match calculation are given in Fig.~\ref{match}, for systems with parameters $p\in[10,60]M$ and $e\in[0.7,0.9]$, with the remaining parameters the same as those used in Fig.~\ref{snr}. The match is always greater than $0.95$ for all systems studied, with only systems at low eccentricities and large semi-latus recta having matches below 0.97. The reason for this is that for such large values of $p$, only the high frequency exponential tail of the numerical waveforms is ``in band" of the LIGO detectors. Meanwhile, the EFB-F2 template has a high frequency cutoff that doesn't accurately track this tail above the frequency in Eq.~\ref{eq:fcut}. As an alternative to the cutoff frequency, one could apply a high frequency filter to cancel out the exponential growth created by the transformation in Eq.~\ref{eq:2f1-id}. Since the systems with low matches also corresponds to systems with low SNR that are likely undetectable, we do not explore this here.

\section{Fisher Analysis}

The draw of having fast to evaluate waveforms is the desire to perform parameter estimation in reasonable amounts of time. Since the EFB-F2 waveforms are analytic, we may employ a Fisher analysis to study the uncertainties of the waveform's parameters without the need of computationally expensive techniques.

\subsection{Single Bursts}
\label{single}
To study the accuracy to which we can measure the parameters of the bursts, we use a Fisher analysis, with the Fisher information matrix given by
\begin{equation}
\Gamma_{ab} = \left(\frac{\partial h}{\partial \lambda^{a}} \Big| \frac{\partial h}{\partial \lambda^{b}}\right)
\end{equation}
where $\lambda^{a}$ are the parameters of the waveform. The variance of the parameters is found by inverting the Fisher matrix, specifically
\begin{equation}
\Delta \lambda^{a} = \left[\Gamma^{-1}\right]_{aa}\,.
\end{equation}
Further, the correlations among parameters are characterized by the correlation coefficients, specifically the off-diagonal components of the inverse Fisher matrix
\begin{equation}
c_{ab} = \frac{\left[\Gamma^{-1}\right]_{ab}}{\sqrt{\left[\Gamma^{-1}\right]_{aa} \left[\Gamma^{-1}\right]_{bb}}}\,.
\end{equation}
From a practical perspective, inverting a numerical matrix can be problematic. To perform the matrix inversion, we use the linear algebra methods in the \texttt{numpy} package of \texttt{python} to perform a singular value decomposition (SVD) of the Fisher matrix. From the SVD, we compute the ratio of the minimum eigenvalue of the Fisher matrix to its maximum eigenvalue. If this ratio is smaller than the numerical accuracy of our computation, then the matrix inversion is badly conditioned. Since we are working at double precision accuracy, we require the ratio to be greater than $10^{-14}$. As a further test of the validity of the numerical inversion, we also compute the product of the Fisher matrix with its numerical inverse, and compare to the identity matrix. Since the procedure for obtaining the inverse is numerical, the off-diagonal components will not necessarily be zero, which gives us a further estimator of numerical error in the inversion. For the cases studies that satisfy the above requirement on the eigenvalues, the greatest value of the off-diagonal components is typically of the order $10^{-6} - 10^{-5}$.

Another practical issue is related to which parameters to include in the Fisher analysis. By studying the EFB-F2 waveform in Eq.~\eqref{eq:efb-f2}, one might expect the ten parameters to be the total mass $M$, the symmetric mass ratio $\eta$, the semi-latus rectum $p_{0}$, the orbital eccentricity $e_{0}$, the time of pericenter passage $t_{p}$, the inclination angle $\iota$, the polarization angle $\beta$, the luminosity distance of the source $D_{L}$, and the two angles characterizing the source's sky locations $(\theta, \phi)$ which enter through the detector response
\begin{equation}
h = F_{+}(\theta, \phi) h_{+} + F_{\times}(\theta,\phi) h_{\times}
\end{equation}
with $(F_{+}, F_{\times})$ the beam pattern functions of the detector, specifically
\begin{align}
F_{+} &= \frac{1}{2} \left(1 + \cos^{2}\theta\right) \cos(2\phi)\,,
\\
F_{\times} &= \frac{1}{2} \cos\theta \sin(2\phi)\,.
\end{align}
In the course of our investigation, we found that including the sky location in the Fisher analysis always results in badly conditioned matrices for inversion. This arises due to the sky location being poorly constrained with single detectors. Further, the total mass and mass ratio $(M, \eta)$ also result in the Fisher matrix being badly conditioned. The reason for this is that there is a degeneracy between the parameters that results in these two quantities not being independently measurable. Such a degeneracy also occurs in the leading PN order quasi-circular TaylorF2 waveforms, specifically the measurable parameter is the binary's chirp mass ${\cal{M}} = M \eta^{3/5}$. In the case of the EFB-F2 waveform, there is an additional degeneracy between the semi-latus rectum and the total mass, such that the measurable quantity is the orbital radius of curvature defined as ${\cal{P}} = p^{3/2}/m^{1/2}$. Note that these two degeneracies were previously found in the leading PN order burst model developed in~\cite{Loutrel:2014vja}. Thus, for our Fisher analysis, the parameters are $\lambda^{a} = \left(\ln{\cal{M}}, \ln{\cal{P}}_{0}, e_{0}, t_{p}, \ln D_{L}, \cos\iota, \beta\right)$. Note that we use the natural logarithm of several parameters, as well as the cosine of the inclination angle in the analysis. Generally, this causes the Fisher matrix to be better conditioned for numerical inversion.

\begin{figure*}
	\begin{subfigure}
		\centering
		\includegraphics[scale=0.3]{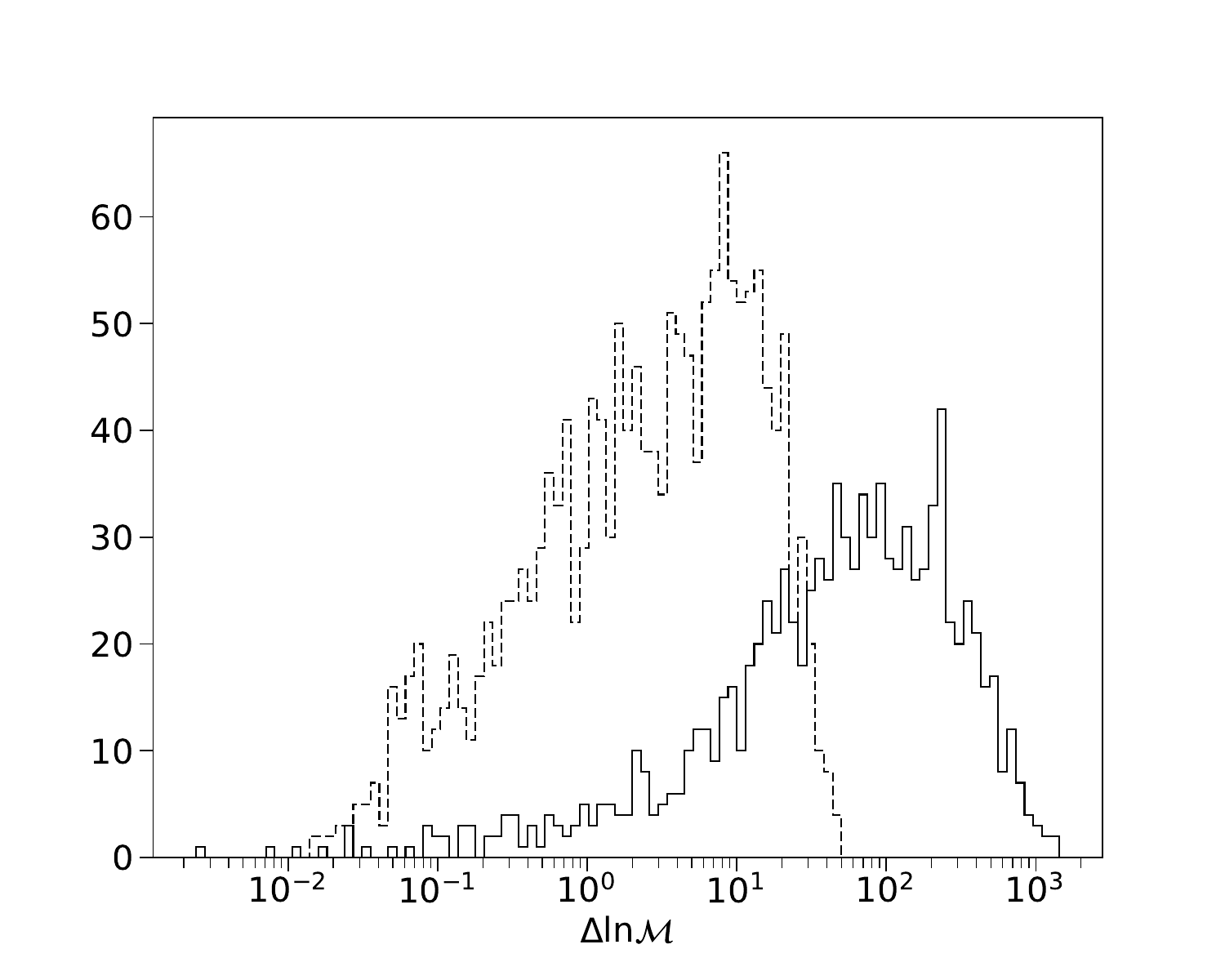}
		\includegraphics[scale=0.3]{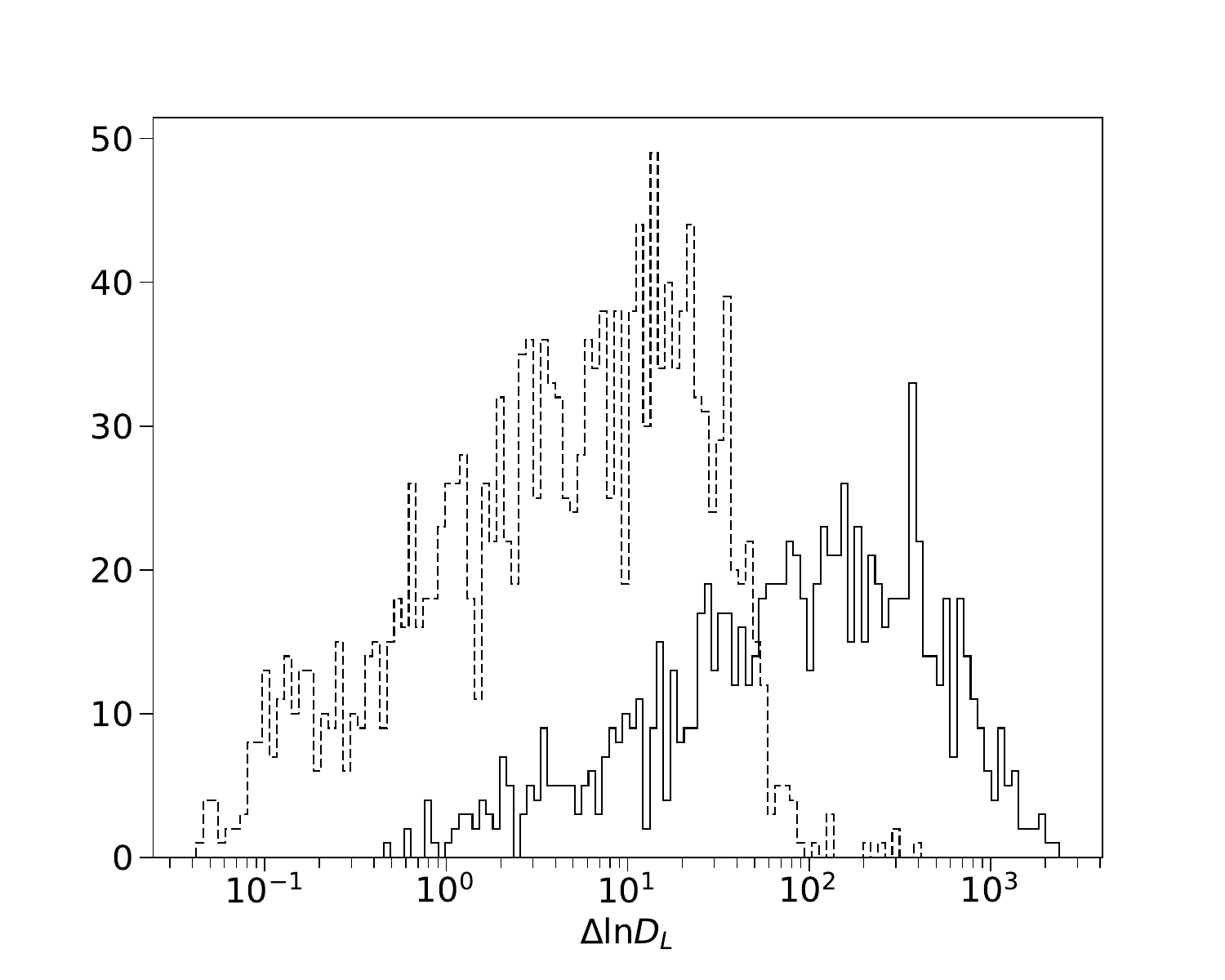}
	\end{subfigure}
	
	\begin{subfigure}
		\centering
		\includegraphics[scale=0.3]{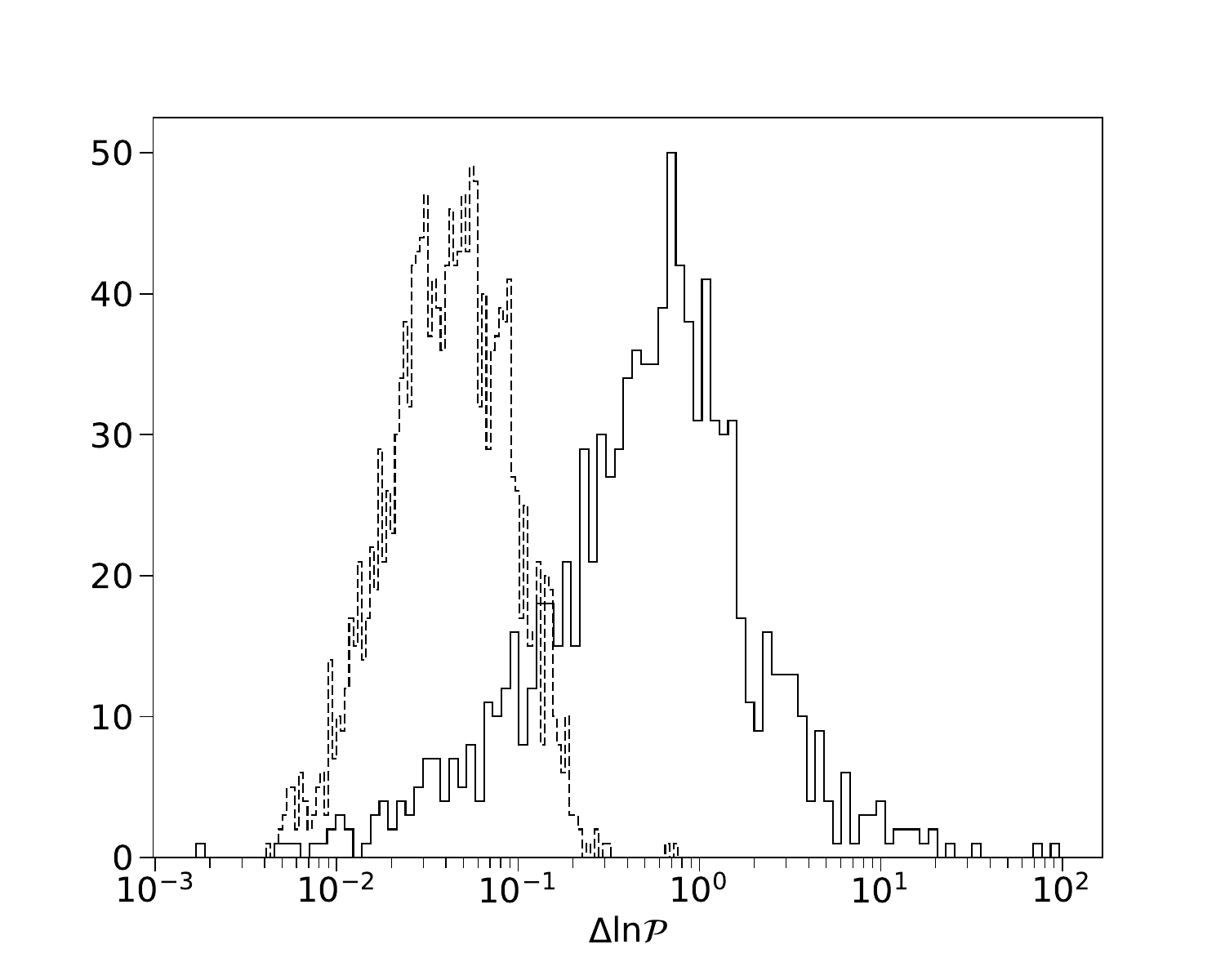}
		\includegraphics[scale=0.3]{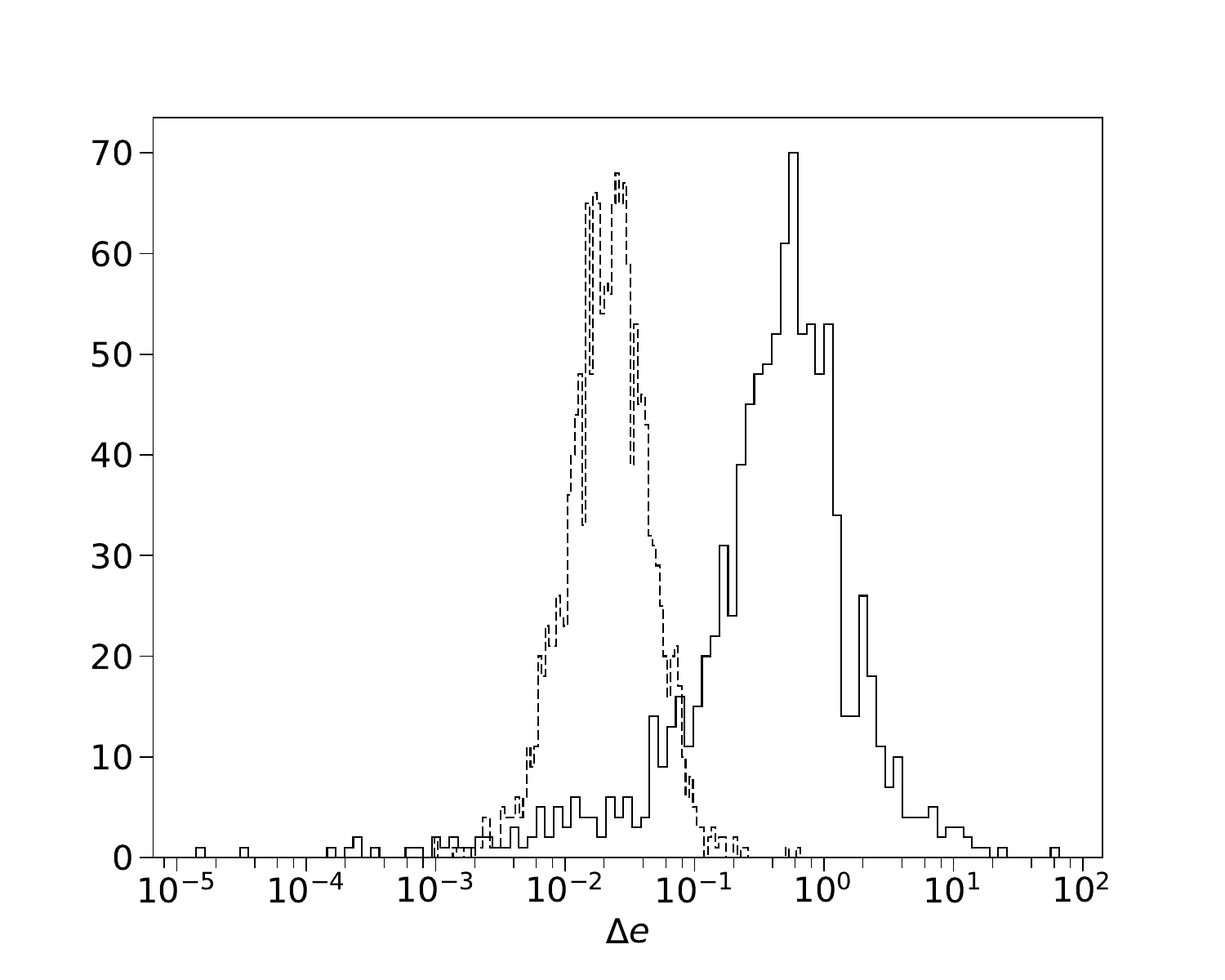}
	\end{subfigure}
	
	\begin{subfigure}
		\centering
		\includegraphics[scale=0.3]{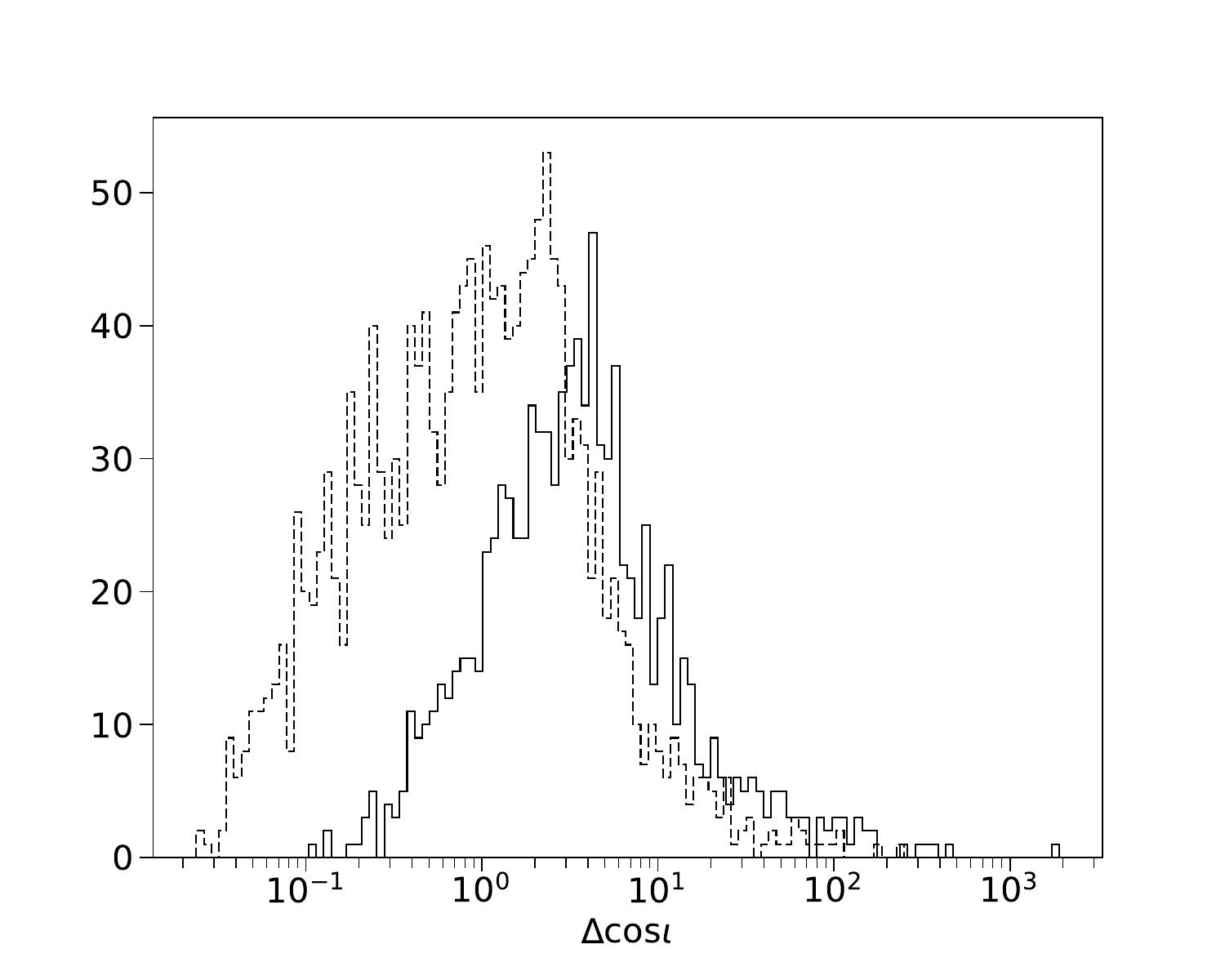}
		\includegraphics[scale=0.3]{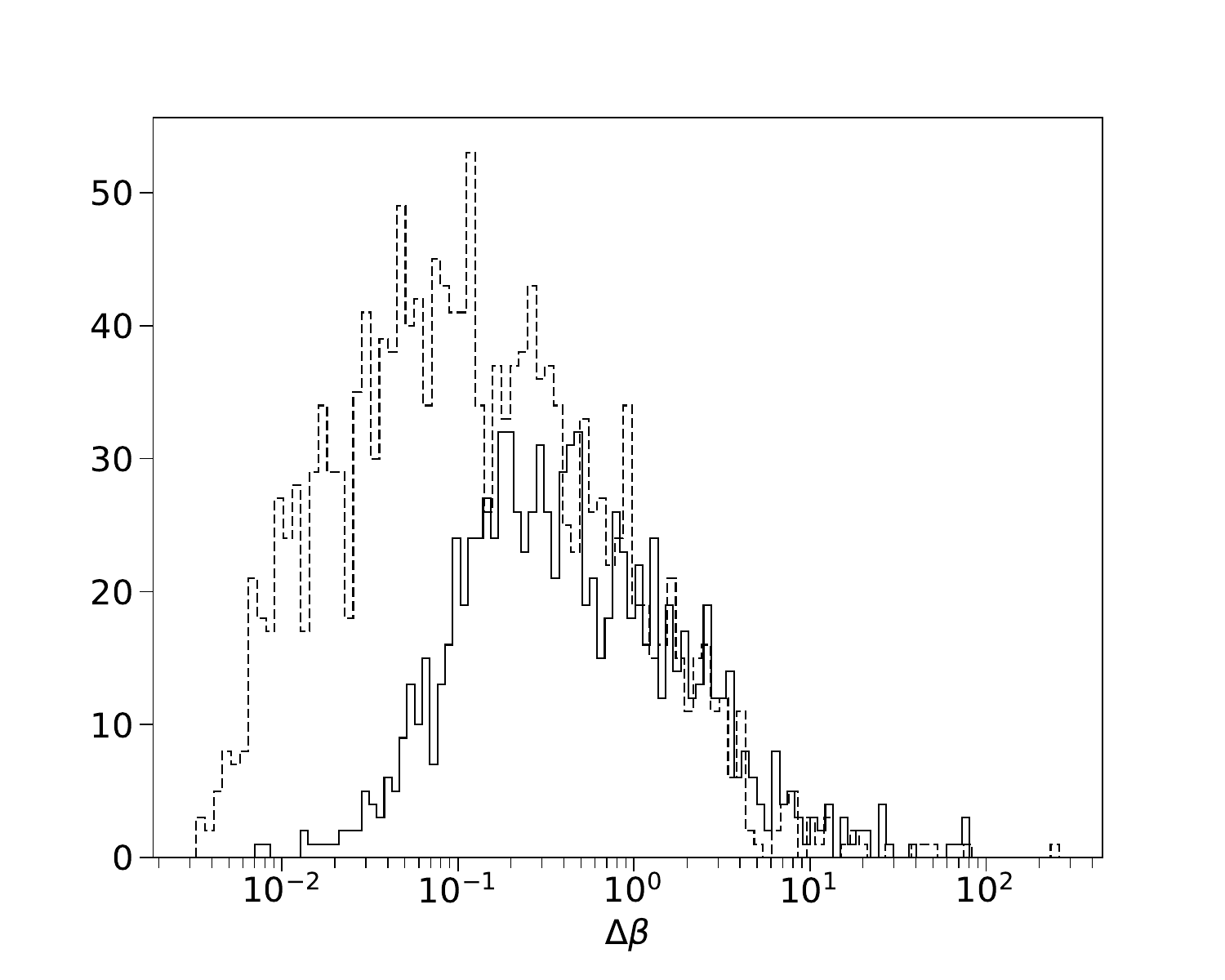}
	\end{subfigure}
	\label{fisher-plot}
	\caption{Histograms of the statistical uncertainty in the parameters of the EFB-F2 model: chirp mass ${\cal{M}}$ (top left), luminosity distance $D_{L}$ (top right), orbital radius of curvature ${\cal{P}}$ (middle left), orbital eccentricity $e$ (middle right), inclination angle $\iota$ (bottom left), and polarization angle $\beta$ (bottom right). The solid histogram gives the distribution of uncertainties for single bursts events, while the dashed histogram includes the following burst for the same systems as predicted by the leading PN order timing model.}
\end{figure*}

For our analysis, we fix the masses of the binary to be $m_{1} = 10 M_{\odot} = m_{2}$, and the luminosity distance $D_{L} = 10$ Mpc. We then generate two thousand systems by randomly selecting the remaining parameters from the ranges $p \in [10,50] M$, $e\in [0.7,0.999]$, $(\iota, \theta) \in [0,\pi]$, and $(\beta,\phi) \in [0,2\pi]$. For each system, we calculate the Fisher matrix for single bursts given by the EFB-F2 waveform and compute the ratio of the minimum eigenvalue to the maximum eigenvalue, as well as the SNR of the source, given by $\rho = (h|h)^{1/2}$. We then require that the ratio of eigenvalues to be greater than $10^{-14}$ and the SNR to be greater than 10. Of the two thousand systems, 993 meet these requirements for a single burst. The results for these systems are given by the solid histogram in Fig.~\ref{fisher-plot}. 

Generally, all of the parameters have large uncertainties, specifically $\Delta \lambda^{a} > 1$. For the chirp mass and luminosity distance, the distributions peak at approximately $10^{2}$, with only a small subset having $\Delta \ln {\cal{M}} < 1$ or $\Delta \ln D_{L} < 1$. On the other hand, the distributions in $\Delta \ln {\cal{P}}$ and $\Delta e$ peak around one, meaning approximately half of the systems studied have uncertainties below one hundred percent. A similar behavior is found in the polarization angle $\beta$, while most of the systems have $\Delta \cos\iota >1$. 

\begin{figure*}
	\begin{subfigure}
		\centering
		\includegraphics[scale=0.3]{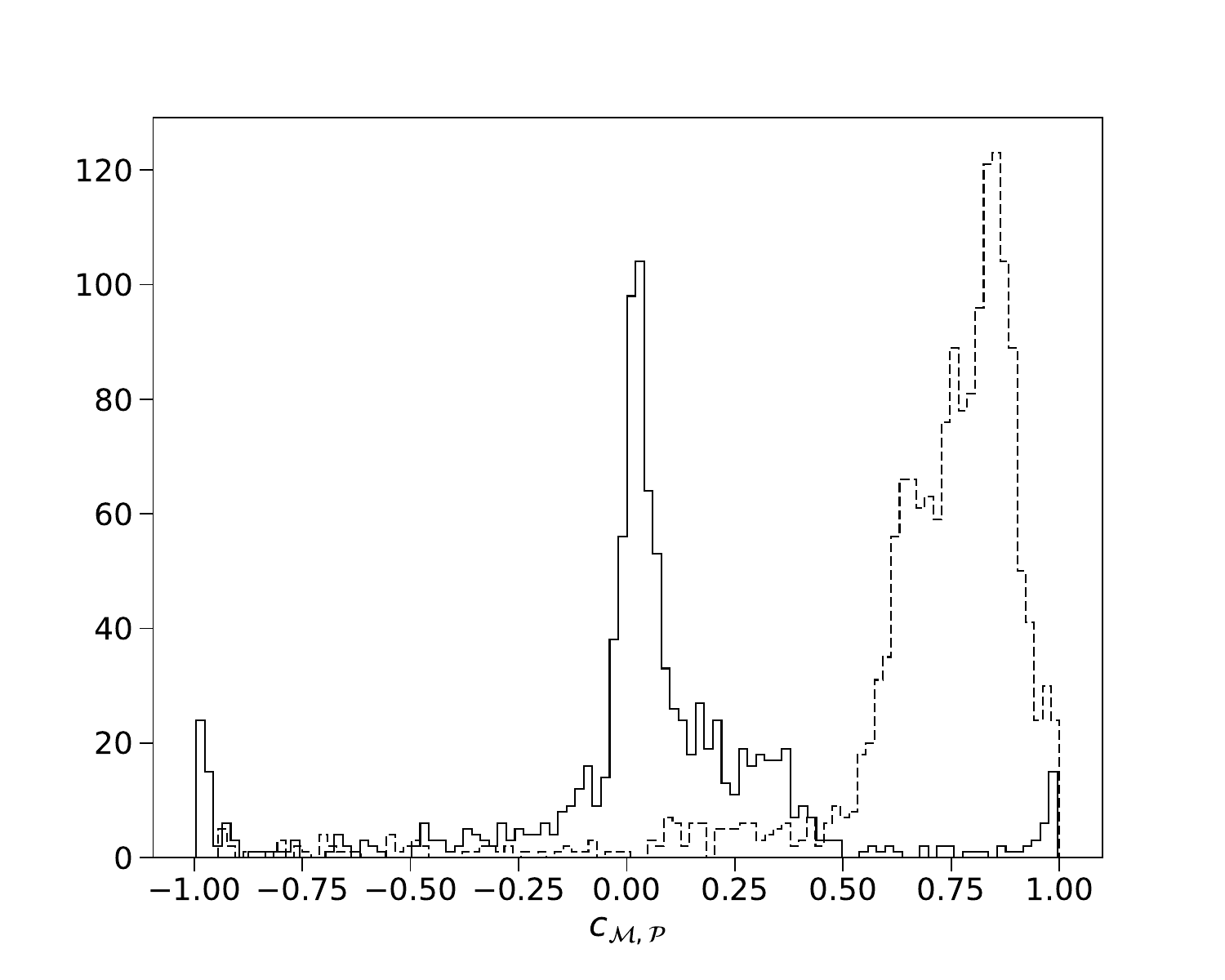}
		\includegraphics[scale=0.3]{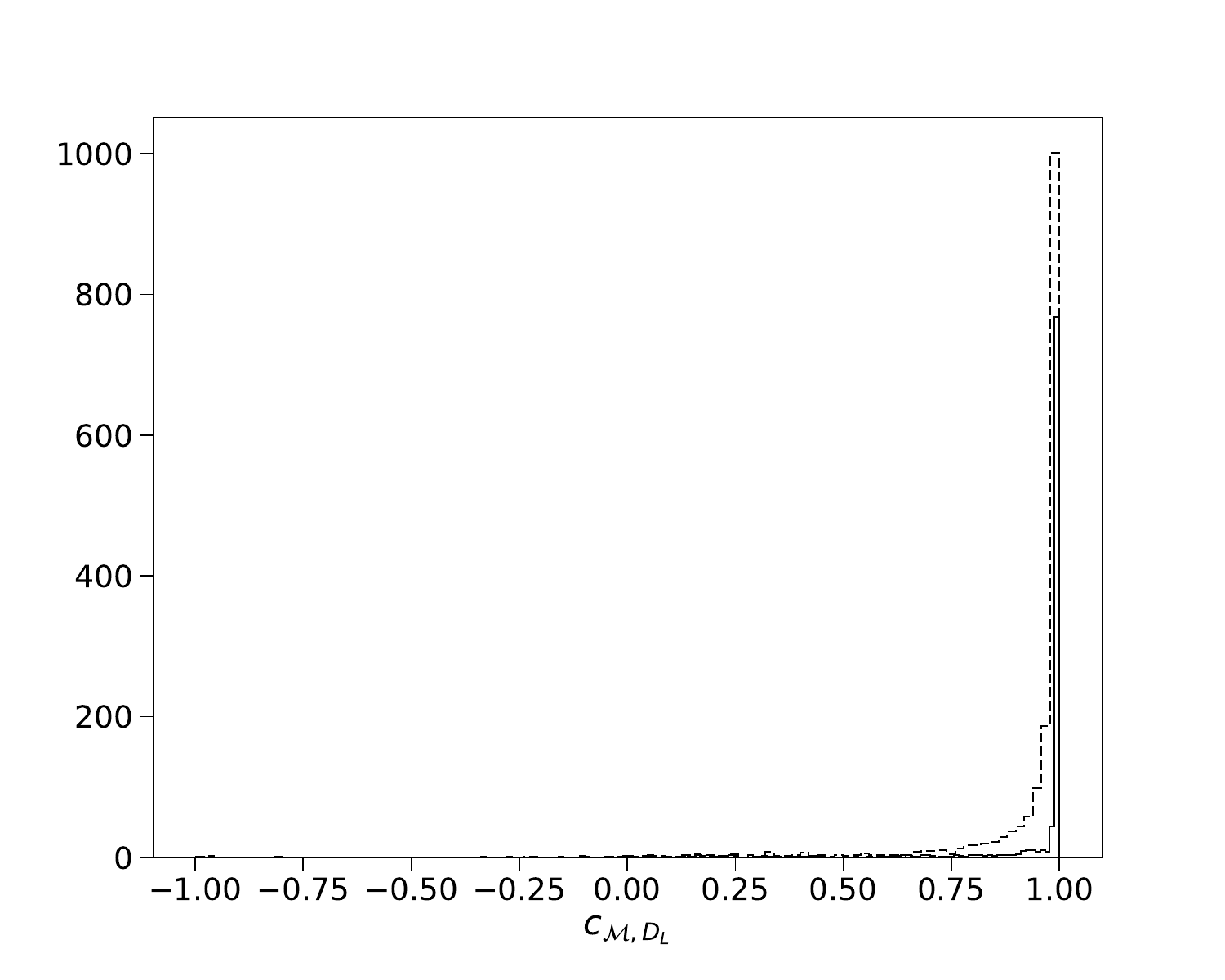}
	\end{subfigure}
	
	\begin{subfigure}
		\centering
		\includegraphics[scale=0.3]{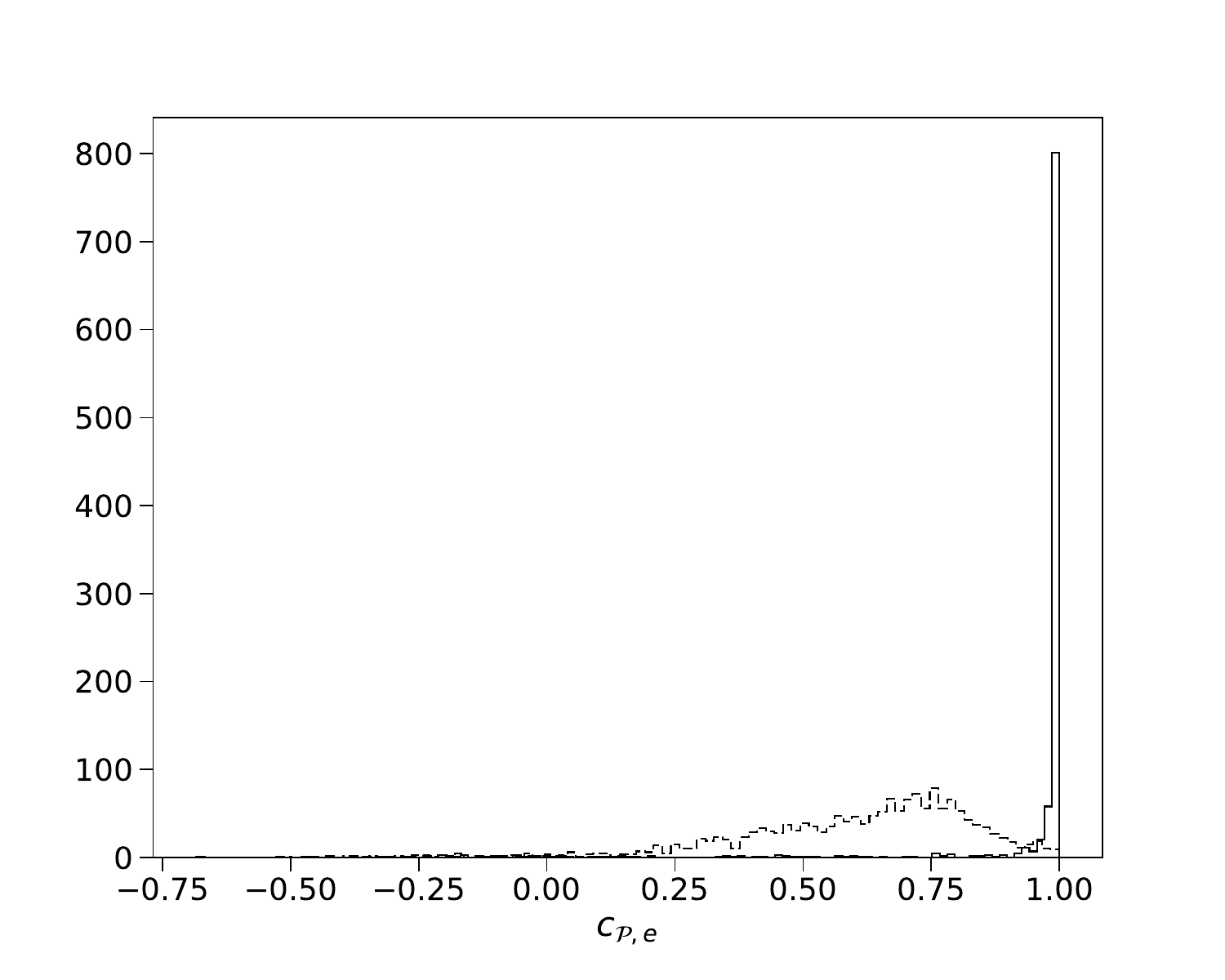}
		\includegraphics[scale=0.3]{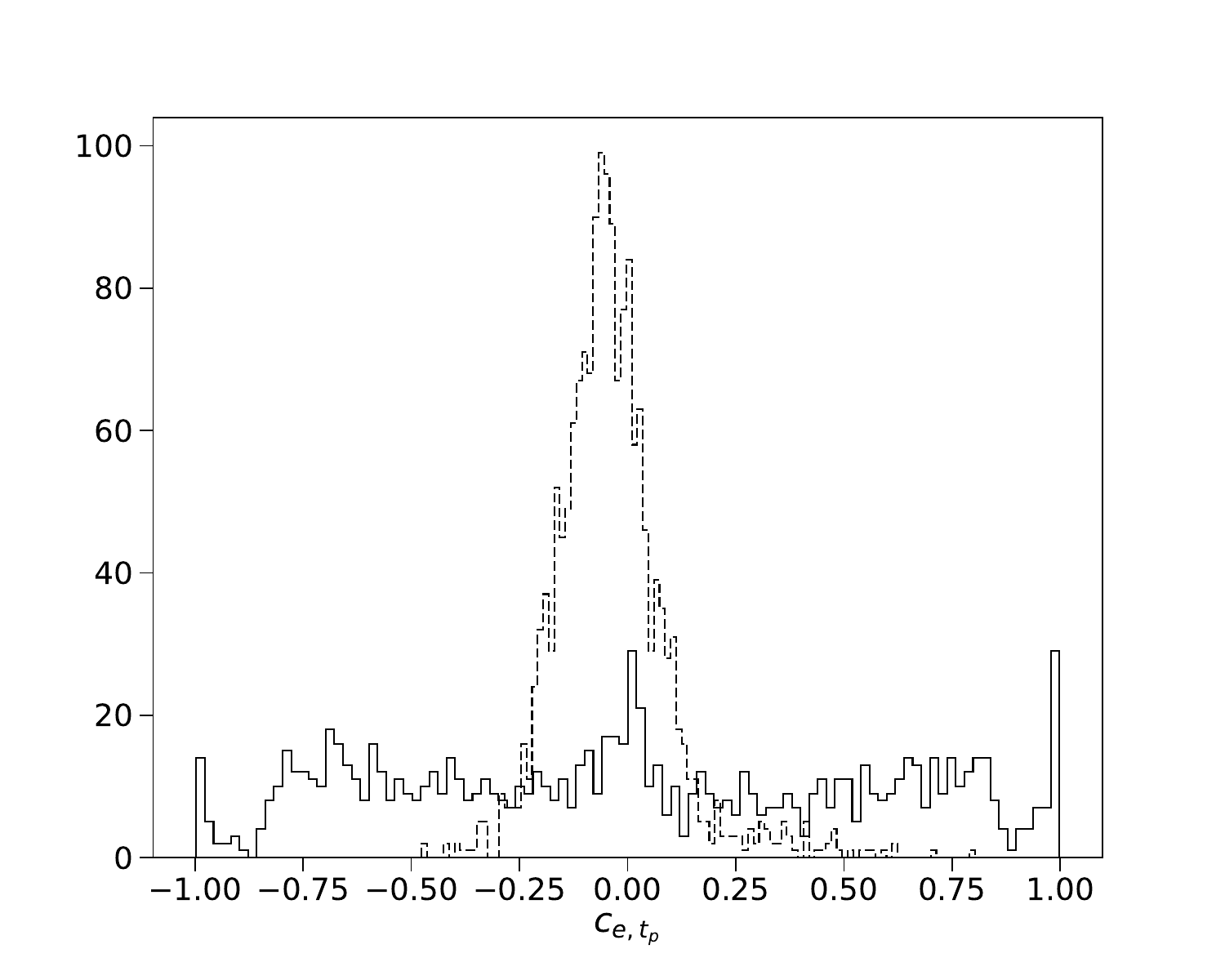}
	\end{subfigure}
	
	\begin{subfigure}
		\centering
		\includegraphics[scale=0.3]{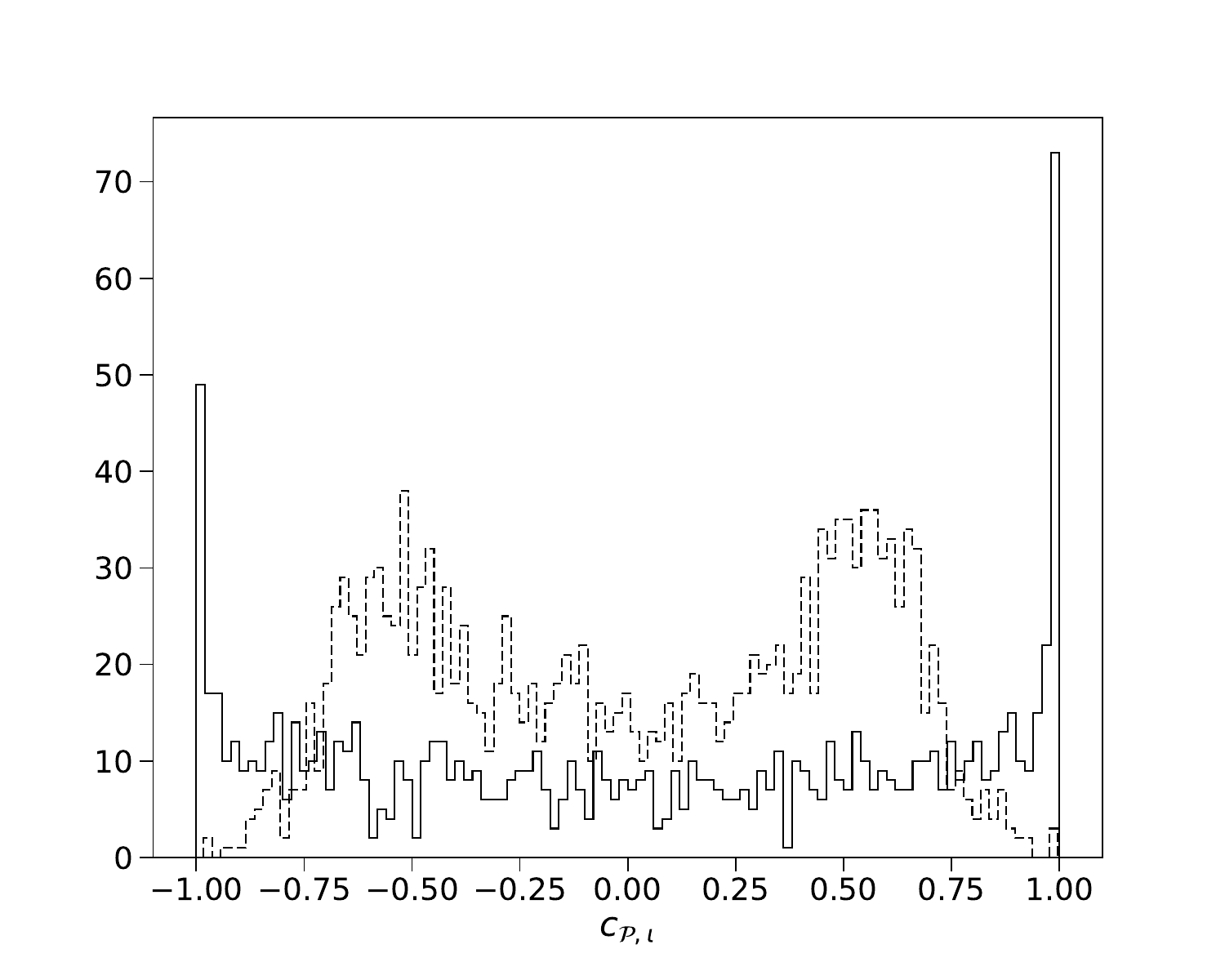}
		\includegraphics[scale=0.3]{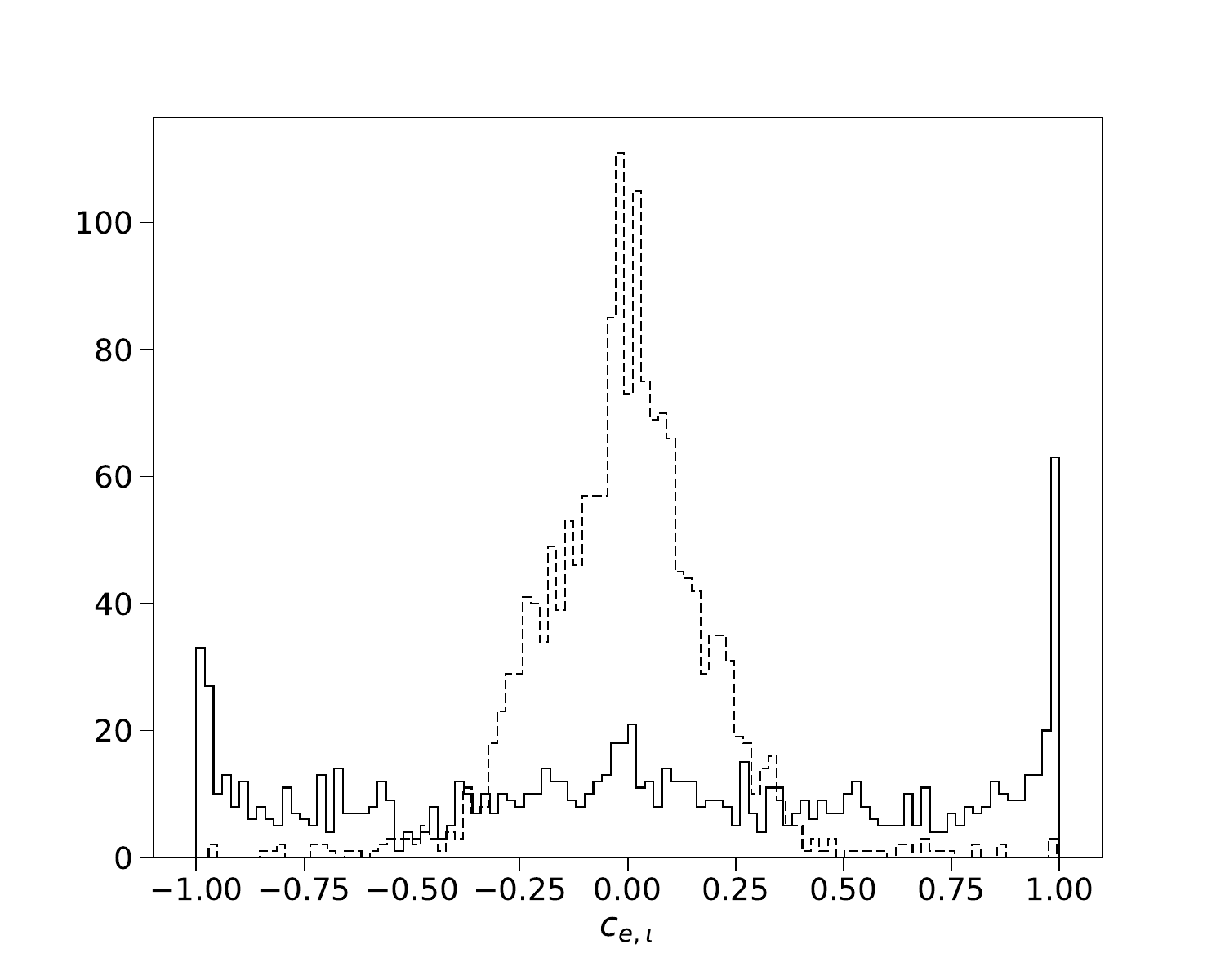}
	\end{subfigure}
	\label{corr-plot}
	\caption{Histograms of the correlation coefficients $c_{{\cal{M}},{\cal{P}}}$ (top left), $c_{{\cal{M}}, D_{L}}$ (top right), $c_{{\cal{P}},e}$ (middle left), $c_{e,t_{p}}$ (middle right), $c_{{\cal{P}},\iota}$ (bottom left), and $c_{e,\iota}$ (bottom right). The solid histogram displays the values for single bursts, while the dashed histogram includes the following burst as predicted by Eqs.~\eqref{eq:pnext}-\eqref{eq:tnext}.}
\end{figure*}

We also plot the distribution of several of the correlation coefficients in solid histograms of Fig.~\ref{corr-plot}. The correlation coefficients generally range from $[-1,1]$, with values at the extremes corresponding to parameters that are completely correlated, or anti-correlated in the case of negative values. In many of the systems studied, there are strong correlations among many of the parameters indicated by peaks in the distributions near $1$ and $-1$. For example, in the top right plot, nearly all of the systems have $c_{{\cal{M}}, D_{L}} \approx 1$, which results from the amplitude of the waveform scaling as ${\cal{M}}/D_{L}$. Similar behavior is found in $c_{{\cal{P}}, e}$ (middle left), which likely results from the scaling of $\chi_{\rm orb}$ and $F_{\rm rr}$ with these parameters. Other parameters have much broader distributions, but still display peaks at extreme values of the correlation coefficients, as can be seen from both $c_{{\cal{P}},\iota}$ (bottom left) and $c_{e,\iota}$ (bottom right). These correlations result from ``face-on" systems having higher SNR, which decreases the uncertainty in parameters.


\begin{figure*}
	\centering
	\includegraphics[scale=0.3]{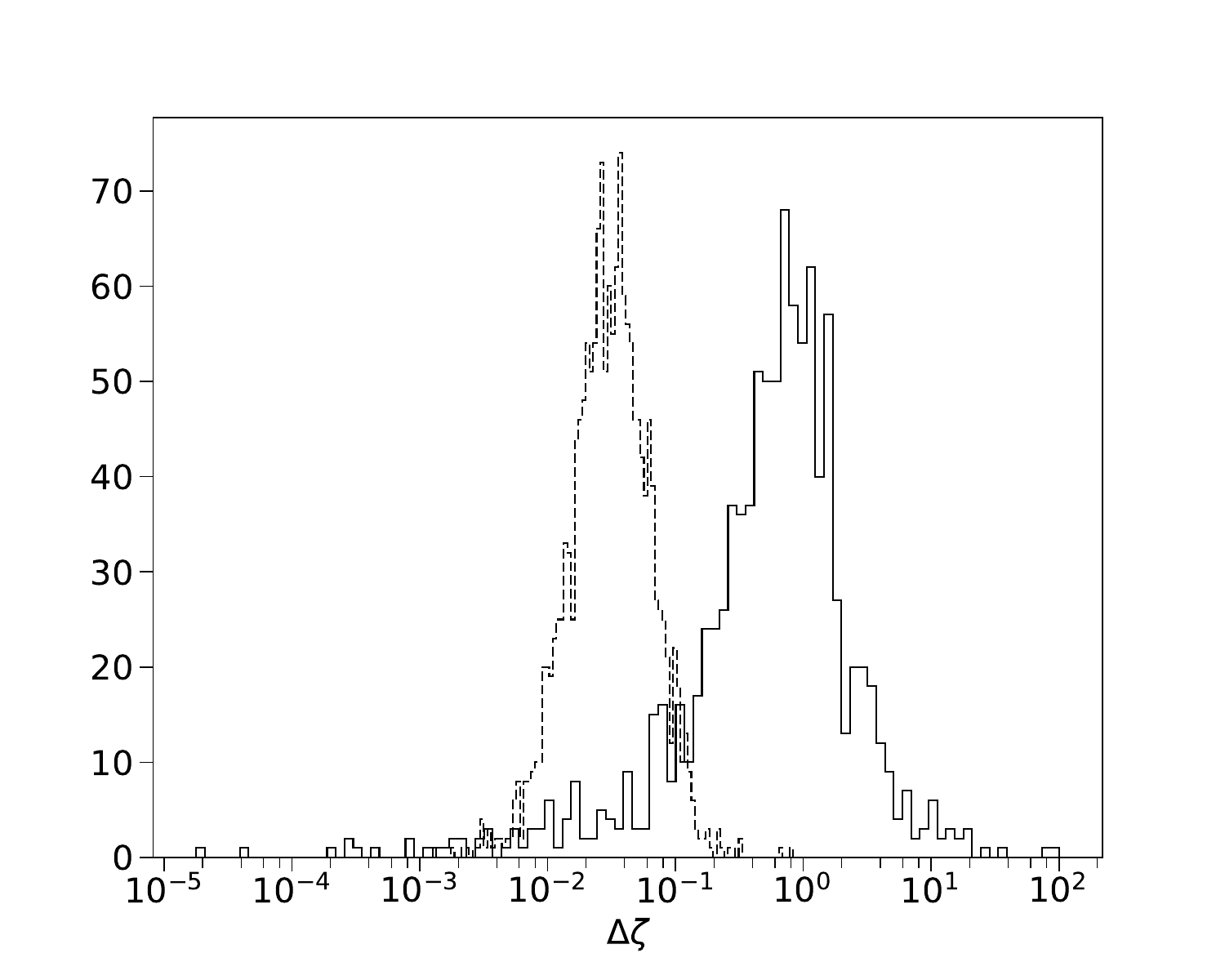}
	\includegraphics[scale=0.3]{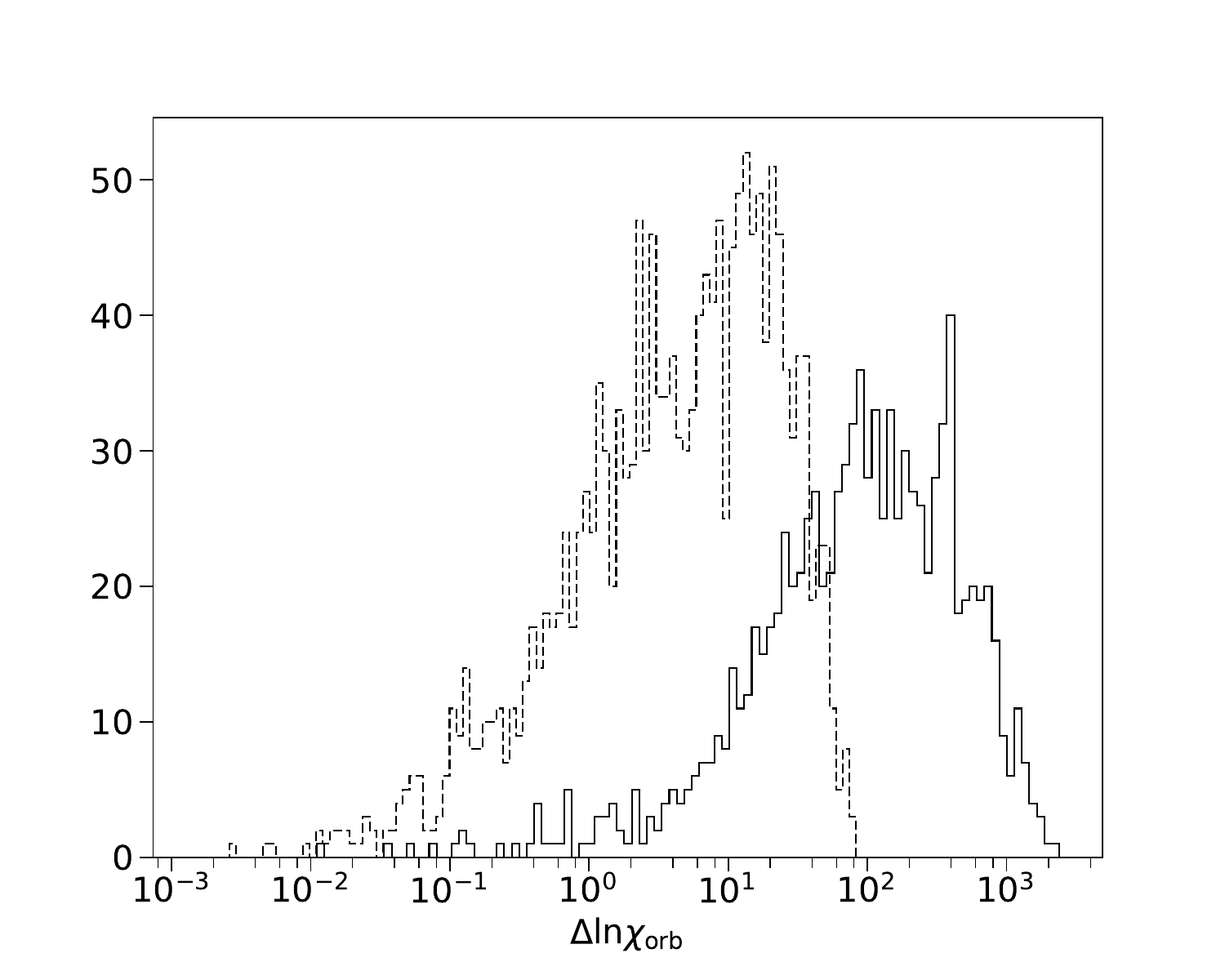}
	\caption{\label{fisher-alt} Histograms of the uncertainty in the alternative EFB-F2 waveform parameters $\zeta$ (left) and $\chi_{\rm orb}$ (right), calculated via Eqs.~\eqref{eq:dzeta}-\eqref{eq:dFrr}, and for both single bursts (solid) and two bursts (dashed). The systems are the same as those used in Fig.~\ref{fisher-plot}. The uncertainties do not show significant improvement from the physical parameters $({\cal{M}}, {\cal{P}}, e)$. }
\end{figure*}

The large uncertainties in the parameters may seem counterintuitive when one compares these results to those of Fisher calculations of quasi-circular binaries using TaylorF2 waveforms. For example, for BBHs, the uncertainty for the chirp mass from those models is typically $\Delta \ln {\cal{M}} \sim 10^{-6}-10^{-5}$~\cite{cutlerflanagan}. Why is it that the EFB-F2 burst waveforms result is significantly larger uncertainties? One possibility is that there may be additional degeneracies, beyond those of the chirp mass ${\cal{M}}$. and radius of curvature ${\cal{P}}$, that limit our ability to make accurate measurements of the system's parameters. In fact, the EFB-F2 waveform polarizations in Eq.~\eqref{eq:efb-f2} are partially written implicitly in terms $({\cal{M}}, {\cal{P}}_{0}, e_{0})$ through $(F_{\rm rr}, \chi_{\rm orb}, \zeta_{0})$. One could then argue that it would be better to choose these parameters for the Fisher analysis rather than the physical parameters $({\cal{M}}, {\cal{P}}_{0}, e_{0})$. This is similar to what is commonly done with quasi-circular TaylorF2 waveforms, which may be written with an overall amplitude coefficient ${\cal{A}}$, and a Fisher analysis is performed with respect to this rather than the luminosity distance. To determine if using these new parameters improves parameter estimation, we compute the uncertainties using error propagation and properly taking into account the correlations among parameters, specifically
\allowdisplaybreaks[4]
\begin{widetext}
\begin{align}
\label{eq:dzeta}
\Delta \zeta &= \frac{1 - e^{2} + \sqrt{1-e^{2}}}{\zeta^{1/2} e \left(1 + \sqrt{1-e^{2}}\right)} \Delta e\,,
\\
\left(\frac{\Delta \chi_{\rm orb}}{\chi_{\rm orb}}\right)^{2}&= \left(\frac{5}{3}\frac{\Delta {\cal{M}}}{{\cal{M}}}\right)^{2} + \left(\frac{5}{3} \frac{\Delta {\cal{P}}}{{\cal{P}}}\right)^{2} + f_{1}(e)^{2} \Delta e^{2}
- 2 c_{{\cal{M}}, {\cal{P}}} \left(\frac{5}{3} \frac{\Delta {\cal{M}}}{{\cal{M}}}\right) \left(\frac{5}{3} \frac{\Delta {\cal{P}}}{{\cal{P}}}\right) 
+ 2 c_{{\cal{M}}, e} f_{1}(e) \left(\frac{5}{3} \frac{\Delta {\cal{M}}}{{\cal{M}}}\right) \Delta e
\nn \\
&- 2 c_{{\cal{P}}, e} f_{1}(e) \left(\frac{5}{3} \frac{\Delta {\cal{P}}}{{\cal{P}}}\right)  \Delta e\,,
\\
\label{eq:dFrr}
\left(\frac{\Delta F_{\rm rr}}{F_{\rm rr}}\right)^{2} &= \left(\frac{5}{3}\frac{\Delta {\cal{M}}}{{\cal{M}}}\right)^{2} + \left(\frac{8}{3} \frac{\Delta {\cal{P}}}{{\cal{P}}}\right)^{2} + f_{2}(e)^{2} \Delta e^{2} - 2 c_{{\cal{M}}, {\cal{P}}} \left(\frac{5}{3}\frac{\Delta {\cal{M}}}{{\cal{M}}}\right) \left(\frac{8}{3} \frac{\Delta {\cal{P}}}{{\cal{P}}}\right) + 2 c_{{\cal{M}}, e} f_{2}(e) \left(\frac{5}{3}\frac{\Delta {\cal{M}}}{{\cal{M}}}\right) \Delta e
\nn \\
& - 2 c_{{\cal{P}}, e} f_{2}(e) \left(\frac{8}{3} \frac{\Delta {\cal{P}}}{{\cal{P}}}\right) \Delta e\,,
\end{align}
\end{widetext}
with
\begin{align}
f_{1}(e) &= \frac{776 e + 148 e^{3} - 74 e^{5}}{(1 - e^{2}) (96 + 292 e^{2} + 37 e^{4})}\,,
\\
f_{2}(e) &= \frac{488 e - 728 e^{3} - 185 e^{5}}{(1 - e^{2}) (96 + 292 e^{2} + 37 e^{4})}\,.
\end{align}
We plot histograms of the uncertainties in $\zeta$ and $\chi_{\rm orb}$ in Fig.~\ref{fisher-alt}. The uncertainties do not show improvement over the uncertainties in $({\cal{M}}, {\cal{P}}, e)$, with the uncertainty in $\chi_{\rm orb}$ being dominated by the uncertainty in ${\cal{M}}$.  Thus, a change in parameters $({\cal{M}}, {\cal{P}}, e) \rightarrow (F_{\rm rr}, \chi_{\rm orb}, \zeta)$ does not change the results of the Fisher analysis, suggesting that the large uncertainties are not the results of additional degeneracies.

What, then, is causing the large uncertainties in the waveform's parameters? When considering parameter estimation with single bursts, one has to remember that a single burst corresponds to a single pericenter passage, and not a complete orbital cycle. It is only through the accumulation of the GW signal across many orbital cycles that placing such stringent limits on parameters is possible. In fact, stellar mass quasi-circular binaries typically evolve through several hundred to greater than one thousand orbital cycles, depending on the mass, as they coalesce through the frequency band of ground based detectors. Now, the question is, how does parameter estimation improve as we considering more orbital cycles, and thus more bursts, from a highly eccentric system.

%

\subsection{Multiple Bursts}
\label{multi}
In the previous section, our Fisher analysis focused on single bursts. However, the EFB waveforms are designed to model bound systems, which will emit multiple bursts as they inspiral. It is thus instructive to study how parameter estimation improves with the number of bursts in the inspiral. All EFB waveforms are characterized by three parameters that change from one orbit to the next, namely $(p, e, t_{p})$, or alternatively $({\cal{P}}, e, t_{p})$. Models that calculate iteratively how these parameters depend on those of the previous burst have been developed in~\cite{Loutrel:2014vja, Loutrel:2019kky}, with the leading PN order model given by
\allowdisplaybreaks[4]
\begin{align}
\label{eq:pnext}
{\cal{P}}_{I} &= {\cal{P}}_{I-1} \left[1 - \frac{192 \pi}{5} \left(\frac{{\cal{M}}}{{\cal{P}}_{I-1}}\right)^{5/3} \left(1 + \frac{7}{8} e_{I-1}^{2}\right) \right]\,,
\\
\label{eq:enext}
e_{I} &= e_{I-1} \left[1 - \frac{604\pi}{15} \left(\frac{{\cal{M}}}{{\cal{P}}_{I-1}}\right)^{5/3} \left(1 + \frac{121}{304} e_{I-1}^{2}\right)\right]\,,
\\
\label{eq:tnext}
t_{p,I} &= t_{p,I-1} + \frac{2\pi {\cal{P}}_{I-1}}{(1-e_{I-1}^{2})^{3/2}} 
\nn \\
&\times
\left[1 - \frac{96\pi}{5} \left(\frac{{\cal{M}}}{{\cal{P}}_{I-1}}\right)^{5/3} \left(\frac{1 + \frac{73}{24} e_{I-1}^{2} + \frac{37}{96} e_{I-1}^{4}}{1 - e_{I-1}^{2}}\right)\right]\,.
\end{align}
While each EFB waveform is parameterized by dependent parameters $({\cal{P}}_{I}, e_{I}, t_{p,I})$, the full sequence is only parameterized by the independent parameters $({\cal{P}}_{0}, e_{0}, t_{p,0})$. Thus, caution must be taken when analytically computing the derivatives of the waveforms.

For convenience, we split the parameters $\lambda^{a}$ into two sets, namely extrinsic parameters $\nu^{a} = (D_{L}, \cos\iota, \beta)$, and intrinsic parameters $\mu^{a}_{I} = ({\cal{M}}, {\cal{P}}_{I}, e_{I}, t_{p,I})$. The full waveform is given by the sum of EFB-F2 waveforms, specifically
\begin{equation}
h = \sum_{n=0}^{N} h_{\rm EFB-F2}\left(\mu^{a}_{n}, \nu^{a}\right) = \sum_{n=0}^{N} h_{n}
\end{equation}
The observable parameters are $\nu^{a}$ and $\mu_{0}^{a}$. Since $\nu^{a}$ do not vary from one burst to the next, the derivative is trivially given by
\begin{equation}
\frac{\partial h}{\partial \nu^{a}} = \sum_{n=0}^{N} \frac{\partial h_{n}}{\partial \nu^{a}}\,.
\end{equation}
The derivative with respect to $\mu^{a}_{0}$ is more complicated and requires repeated application of the chain rule, specifically
\begin{equation}
\frac{\partial h}{\partial \mu_{0}^{a}} = \sum_{n=0}^{N} \frac{\partial h_{n}}{\partial \mu_{0}^{a}} + \sum_{n=0}^{N} \left(\prod_{i=1}^{n} \frac{\partial \mu^{a_{i}}_{i}}{\partial \mu^{a_{i-1}}_{i-1}}\right) \frac{\partial h_{n}}{\partial \mu^{a_{n}}_{n}}\,.
\end{equation}
The first term in the above expression is required since the waveform depends on ${\cal{M}}$ not just through $\mu^{a}_{n}$. The second term accounts for the dependence $\mu^{a}_{n}(\mu_{0}^{b})$ through repeated application of the Jacobian $\partial \mu_{i}^{a}/\partial \mu_{i-1}^{b}$. Note that here we make use of the Einstein summation convention, so repeated indices must be summed over.

With the expression for the derivatives now in hand, we repeat the Fisher analysis for the two thousand systems generated in the previous section, now including the second burst in the sequence with the timing model in Eqs.~\eqref{eq:pnext}-\eqref{eq:tnext}. We once again require each of the two burst waveforms to meet the previous requirements of having an SNR$ > 10$, and the ratio of the eigenvalues of the Fisher matrix to be greater than $10^{-14}$. Of the two thousand systems, 1682 now meet the requirements. The results of this computation are given by the dashed histogram in Fig.~\ref{fisher-plot}. The improvement on uncertainties of the parameters depends on the exact parameters of the system, but are typically improved by a factor of 2 at the least, and by more than an order of magnitude at most. It is thus clear that, in order to perform accurate parameter estimation with these waveforms, one will need more than a single burst and an accurate timing model to characterize the burst sequence.

We also plot the new correlation coefficients in the dashed histograms of Fig.~\ref{corr-plot}. The inclusion of an additional burst acts to break some of the correlations that were present for the single burst case. For example, the correlation between the radius of curvature ${\cal{P}}$ and eccentricity $e$ now peaks at $c_{{\cal{P}}, e} \approx 0.75$ rather than one, and the systems studied have a broad range of possibles value for the correlation coefficient instead of the distribution peaking strongly around high values. On the other hand, the timing model that characterizes the phase between the bursts also introduces new correlations, as can be seen in $c_{{\cal{M}}, {\cal{P}}}$. The distribution now peaks at high values of the correlation coefficient and are typically in the range $0.5 < c_{{\cal{M}}, {\cal{P}}} < 1$. By studying Eqs.~\eqref{eq:pnext}-\eqref{eq:tnext}, its not difficult to see why. The chirp mass always enters the timing model coupled to the radius of curvature such that the dependence is $({\cal{M}}/{\cal{P}})^{5/3}$, producing the correlation that is observed. Finally, some of the correlations do not change at all, as can be seen from $c_{{\cal{M}}, D_{L}}$. The reason for this is that the amplitude of every new burst scales as ${\cal{M}}/D_{L}$, and thus the original correlation is not broken by including additional bursts.

\begin{figure*}
	\centering
	\includegraphics[scale=0.25]{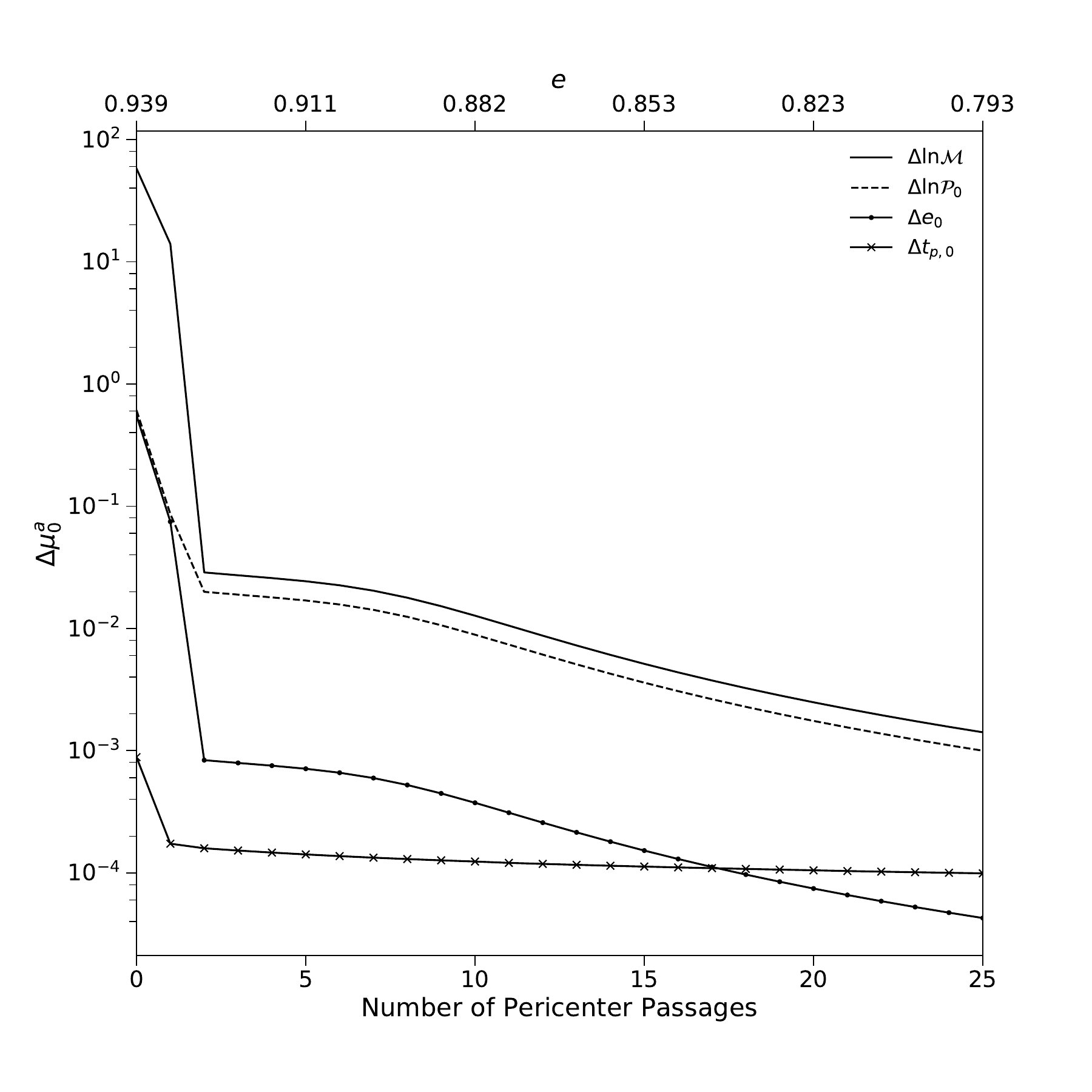}
	\includegraphics[scale=0.25]{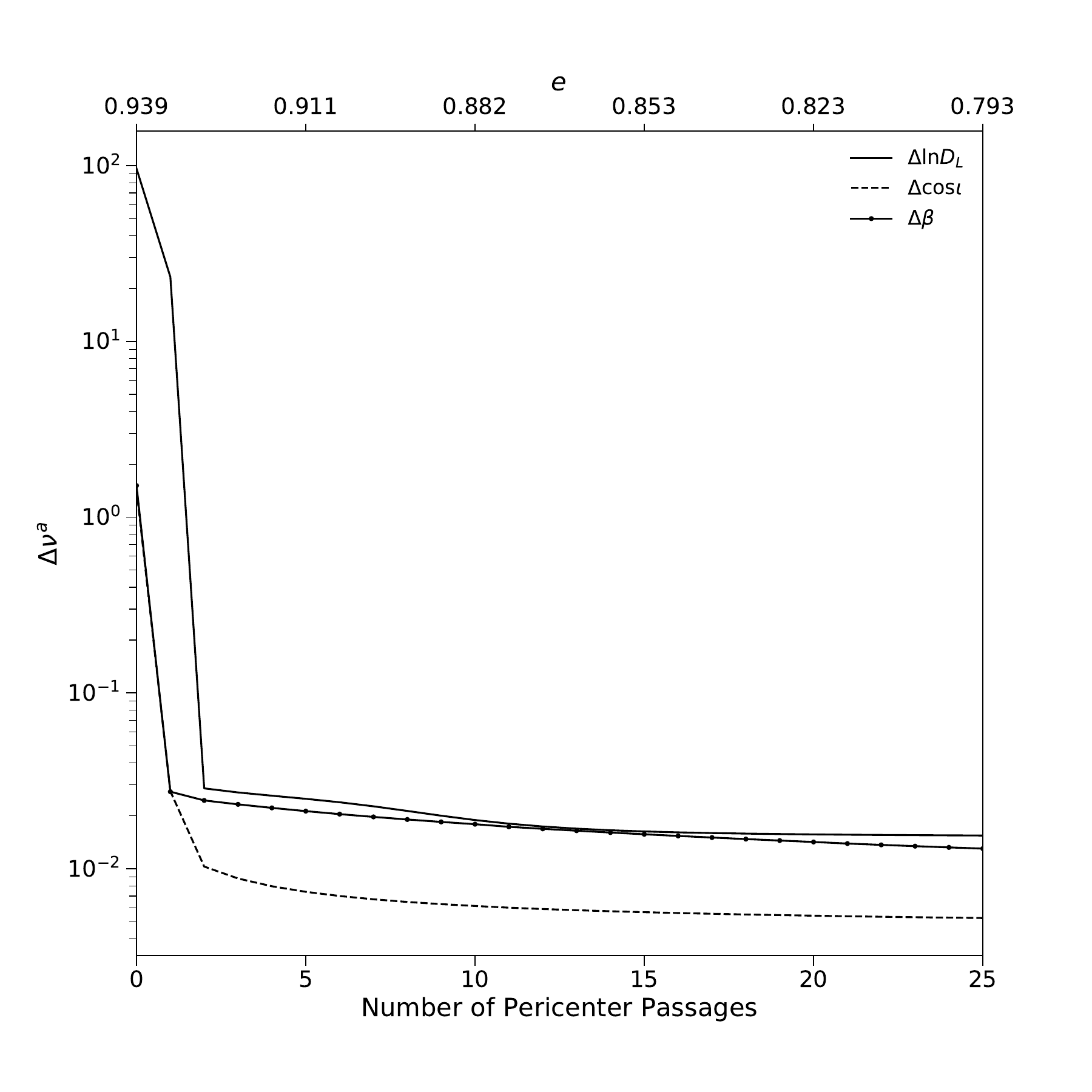}
	\caption{\label{fisher-plot2} Uncertainty in the intrinsic parameters $\mu^{a}_{0} = ({\cal{M}}, {\cal{P}}_{0}, e_{0}, t_{p,0})$ (left) and extrinsic parameters $\nu^{a} = (D_{L}, \iota, \beta)$ (right) as a function of the number of pericenter passages for a single binary system. The burst at each passage is given by a single EFB-F2 waveform using the timing model of Eqs.~\eqref{eq:pnext}-\eqref{eq:tnext}. The first few bursts in the sequence provide a minimum order of magnitude reduction in the uncertainty of the parameters, with the one exception being $t_{p,0}$ which simply corresponds to an overall time shift of the entire sequence. After the third burst, the uncertainties only show slight improvements from one burst to the next, but can still change by orders of magnitude due to the accumulation of power from many bursts.}
\end{figure*}

Does the trend of decreasing uncertainties continue as one considers more bursts within the inspiral sequence? To investigate this, we select one of the systems that passes our SNR and eigenvalue requirements, and repeat the Fisher analysis for a sequence of bursts. The system has parameters $p_{0} = 34.9 M$, $e_{0} = 0.939$, $\cos\iota = -0.231$, $\beta = 3.70$, $\cos\theta = 0.751$, and $\phi = 4.06$. We compute the uncertainties for all of the intrinsic and extrinsic parameters up to the 25th burst after the initial, which we number as zero in the sequence. The results of this analysis are plotted in Fig.~\ref{fisher-plot2}. The left plot shows the uncertainty in the intrinsic parameters $\mu^{a}_{0} = ({\cal{M}}, {\cal{P}}_{0}, e_{0}, t_{p,0})$. The effect of the first two bursts after the initial is to improve the uncertainties by about a factor of 2016 for ${\cal{M}}$, 660 for $e_{0}$, 30 for ${\cal{P}}_{0}$, and five for $t_{p,0}$. After this, the uncertainties no longer show significant improvement from one burst to the next, instead converging to a steady trend of decreasing uncertainty. It is now only through the accumulation of power across many bursts that one obtains orders of magnitude improvements in the uncertainties of the parameters. The same trend is displayed in the right plot of Fig.~\ref{fisher-plot2} for the extrinsic parameters.

As a final point, we comment on why we stop the above analysis after 25 bursts. The top axis of the plots in Fig.~\ref{fisher-plot2} gives the eccentricity parameter at the specific number of pericenter passages on the bottom axis. From this, as well as Eq.~\eqref{eq:enext}, we see that the eccentricity decreases from one passage to the next, a key feature of the leading PN order radiation reaction effects. Eventually, as we consider more pericenter passages, the eccentricity will become so low that the waveform no longer resembles discrete bursts, but instead the continuous waveform of low eccentricity inspirals. In addition, as can be seen from Fig.~\ref{match}, the match between EFB-F2 waveforms and leading PN order numerical waveforms generally decreases with decreasing eccentricity, implying that the EFB-F2 waveforms become less accurate. It is thus necessary to terminate the sequence of EFB-F2 waveforms given by Eqs.~\eqref{eq:pnext}-\eqref{eq:tnext} at some point, and use a moderate to low eccentricity waveform to accurately model the evolution of the system. We terminate the sequence after 25 bursts to ensure we are still comfortably in the range of parameter space where the EFB-F2 waveforms are valid. 
\section{Discussion}
\label{discuss}

We have here performed the first parameter estimation study using analytic waveforms for gravitational bursts from highly eccentric binaries. There are two conclusions that we can draw from our analysis. First, just like quasi-circular binaries at leading PN order, there are degeneracies that prevent the measurement of certain parameters. Specifically, the degeneracies result in the waveform being purely written in terms of the chirp mass ${\cal{M}} =  M \eta^{3/5}$ and the orbital radius of curvature ${\cal{P}} = (p^{3}/M)^{1/2}$, the latter of which is not present in quasi-circular binaries. Second, the Fisher analysis used here suggests that one cannot make accurate measurements of many of the waveform's parameters with single bursts. Further, the Fisher analysis suggests that it is only through the accumulation of phase across multiple bursts that one can perform accurate parameter estimation.

However, are the results of the Fisher analysis performed herein accurate? The ``Holy Grail" of parameter estimation would be to perform Bayesian inference using a Markov Chain Monte Carlo (MCMC) to map the full posterior distribution of the parameters. This method is unfortunately computationally expensive, and faster methods like the Fisher analysis are commonly employed. Unfortunately, there are known issues with the Fisher matrix calculations. It is common for the Fisher analysis to predict greater than one hundred percent uncertainties, while Bayesian inference will give far more reasonable error bounds~\cite{Porter:2015eha}. Further, the Fisher analysis is known to only be valid under certain conditions, specifically when the detector noise is stationary and gaussian, when the SNR is sufficiently (and often unrealistically) large, and when the prior probability distributions on the waveform's parameters can be neglected~\cite{Vallisneri:2007ev}. 

The systems used herein have single burst SNRs in the range $10 \le $SNR$ \lesssim 450$, which already may be unrealistically large for real astrophysical sources. Meanwhile, the SNR of the sequence of bursts grows roughly as $N^{1/2}$, with $N$ the number of bursts. This seems to imply that as we consider more bursts within the inspiral sequence, the Fisher analysis may become a more valid representation of the posterior distribution. The results of Fig.~\ref{fisher-plot2} lend credence to this idea, since the uncertainties converge to a specific monotonically decreasing sequence after only three bursts. Further, as we consider more bursts within the sequence, the ratio of the eigenvalues of the Fisher matrix becomes larger, meaning the Fisher matrix becomes less singular, and thus better conditioned for inversion. Ultimately, the question of whether the results of the Fisher analysis reported herein are accurate needs to be answered by more sophisticated studies of parameter estimation using, for example, Bayesian inference.

\acknowledgements
I would like to thank Leo Stein, whose suggestions and comments on the original EFB-F waveforms became the inspiration behind this study. I would further like to thank Frans Pretorius for providing useful comments on this manuscript. An $\texttt{ipython}$ notebook with the necessary functions to evaluate the waveform model is available upon request. This work was supported by NSF grant PHY-1912171, the Simons Foundation, and the Canadian Institute for Advanced Research (CIFAR).
\appendix
\section{Asymptotic Expansion of Hypergeometric Functions}
\label{2f1-asym}

To obtain the EFB-F2 model, we require an asymptotic expansion of the hypergeometric function, more specifically Eq.~\eqref{eq:2f1-series}, where
\begin{equation}
(a)_{j} = \prod_{k=0}^{j-1} (a - k) = \frac{\Gamma(a+1)}{\Gamma(a-j+1)}\,.
\end{equation}
The Pochhammer symbols can be asymptotically expanded to obtain
\begin{equation}
\left(A - i \frac{\chi}{2}\right)_{j} = e^{\sum_{k=0} g_{k}(A, j) \chi^{-k}}\,,
\end{equation}
where the first few $g_{k}$ coefficients are
\begin{align}
\label{eq:g0}
g_{0} &= \frac{j}{2} \left[2 \log\left(\frac{\chi}{2}\right) - i \pi\right]\,,
\\
g_{1} &= i \left[j^{2} + \left(2A - 1\right) j\right]\,,
\\
\label{eq:g2}
g_{2} &= \frac{1}{3} \left[2 j^{3} + \left(6A - 3\right) j^{2} + \left(6 A^{2} -6 A + 1\right) j\right]\,.
\end{align}
For our purposes, it suffices to truncate the expansion at $k=2$, but it could be extended to higher order if more accuracy is required. Combining this with Eq.~\eqref{eq:2f1-series}, we have
\begin{align}
&\pFq{2}{1}{\frac{l_{1}}{6} - i \frac{\chi}{2}, s - \frac{l_{2}}{6} + i \frac{\chi}{2}}{\frac{l_{1}}{6}-\frac{l_{2}}{6}+1}{\frac{1}{1-X}} 
\nn \\
&\;\;\;\;\;\;\;\; \sim \sum_{j=0}^{\infty} \frac{e^{\sum_{k=0} \left[g_{k}(l_{1}/6, j) + g_{k}^{\dagger}(s-l_{2}/6,j)\right] \chi^{-k}}}{j! (\frac{l_{1}}{6}-\frac{l_{2}}{6}+1)_{j}} \left(\frac{1}{1-X}\right)^{j}\,,
\end{align}
where $\dagger$ corresponds to complex conjugation. In general this series does not have a closed form expression. To re-sum this expression, we regroup terms in the exponential as follows
\begin{equation}
\sum_{k=0} g_{k}(A,j) \chi^{-k} = \sum_{n=1} \tilde{g}_{n}(A,\chi) j^{n}
\end{equation}
where $\tilde{g}_{n}(A,\chi)$ are Laurent series in $\chi$, which may be found by matching terms in Eqs.~\eqref{eq:g0}-\eqref{eq:g2}. For example, the first term is
\begin{align}
\tilde{g}_{1}(A,\chi) &= -\frac{i\pi}{2} + \log\left(\frac{\chi}{2}\right) + \frac{i}{\chi}(2A-1) 
\nn \\
&+ \frac{1}{\chi^{2}}\left(2A^{2}-2A+\frac{1}{3}\right) + {\cal{O}}\left(\frac{1}{\chi^{3}}\right)\,.
\end{align}
We now expand the exponential by first factoring out $\tilde{g}_{1}$ and series expanding the remainder about $\chi \gg 1$, specifically
\begin{align}
e^{\sum_{n=1}\tilde{g}_{n}(A,\chi) j^{n}} &= e^{\tilde{g}_{1}(A,\chi) j} e^{\sum_{n=2}\tilde{g}_{n}(A,\chi) j^{n}}
\nn \\
&= e^{\tilde{g}_{1}(A,\chi)j} \left[1 + \sum_{n=2} \tilde{g}_{n}(A,\chi) j^{n} + {\cal{O}}(\tilde{g}_{n}^{2}) \right]\,.
\end{align}
The reason we may do this is that $\tilde{g}_{n>1} \sim \chi^{-2}$ to leading order. Recombining everything, we are left with
\begin{align}
&\pFq{2}{1}{\frac{l_{1}}{6} - i \frac{\chi}{2}, s - \frac{l_{2}}{6} + i \frac{\chi}{2}}{\frac{l_{1}}{6}-\frac{l_{2}}{6}+1}{\frac{1}{1-X}} 
\nn \\
&\;\;\;\;\;\;\;\; \sim \sum_{j=0}^{\infty} \frac{1}{j! (\frac{l_{1}}{6} - \frac{l_{2}}{6} + 1)_{j}} \left[\frac{e^{{\cal{G}}_{1}(l_{1}, l_{2}, s, \chi)}}{1-X}\right]^{j}
\nn \\
&\;\;\;\;\;\;\;\; \times \left[1 + {\cal{G}}_{2}(l_{1}, l_{2}, s,\chi) j^{2} + {\cal{G}}_{3}(l_{1},l_{2},s,\chi) j^{3} + {\cal{O}}\left(\frac{1}{\chi^{3}}\right)\right]\,,
\end{align}
where ${\cal{G}}_{n}(l_{1}, l_{2}, s, \chi) = \tilde{g}_{n}(l_{1}/6, \chi) + \tilde{g}^{\dagger}_{n}(s-l_{2}/6,\chi)$. This expression can be re-summed to finally obtain
\begin{align}
&\pFq{2}{1}{\frac{l_{1}}{6} - i \frac{\chi}{2}, s - \frac{l_{2}}{6} + i \frac{\chi}{2}}{\frac{l_{1}}{6}-\frac{l_{2}}{6}+1}{\frac{1}{1-X}} 
\nn \\
&\;\;\;\;\;\;\;\; \sim \sum_{n=0}^{2} {\cal{J}}_{n}(l_{1}, l_{2}, s, \chi) J_{\frac{1}{6}(l_{1}-l_{2})+n}\left(\frac{2 e^{{\cal{G}}_{1}/2}}{\sqrt{X-1}}\right)\,,
\end{align}
where $J_{n}(x)$ is the Bessel function of the first kind, and
\allowdisplaybreaks[4]
\begin{align}
{\cal{J}}_{0} &= \left(\frac{e^{{\cal{G}}_{1}/2}}{\sqrt{X-1}}\right)^{(l_{2}-l_{1})/6} \Gamma\left(1 + \frac{l_{1} - l_{2}}{6}\right)\,,
\\
{\cal{J}}_{1} &= \left(\frac{e^{{\cal{G}}_{1}/2}}{\sqrt{X-1}}\right)^{1 + (l_{2}-l_{1})/6} \frac{(1 - X) {\cal{G}}_{2} + (1 - X + e^{{\cal{G}}_{1}}){\cal{G}}_{3}}{(X-1)^{2}[1 + (l_{1}-l_{2})/6]} 
\nn \\
&\times \Gamma\left(2 + \frac{l_{1}-l_{2}}{6}\right)\,,
\\
{\cal{J}}_{2} &= \left(\frac{e^{{\cal{G}}_{1}/2}}{\sqrt{X-1}}\right)^{2+(l_{2}-l_{1})/6} \frac{{\cal{G}}_{2} + [1 + (l_{2} - l_{2})/6]{\cal{G}}_{3}}{(X-1)[1+(l_{1}-l_{2})/6]} 
\nn \\
&\times \Gamma\left(2+\frac{l_{1}-l_{2}}{6}\right)\,.
\end{align}

The last stage of the asymptotic expansion of the hypergeometric functions is to expand the prefactor $\gamma(a,b,c)$, which depend on $\chi$ through the Gamma function. To leading order, the asymptotic expansion of the Gamma function is the well known Stirling formula~\cite{NIST}. To obtain a more accurate representation, we carry out the expansion to second order, obtaining
\begin{align}
\gamma\left(\frac{l_{1}}{6} - i \frac{\chi}{2}, \frac{l_{2}}{6} - i \frac{\chi}{2}, s\right) \sim e^{\pi \chi/2} \tilde{\gamma}_{l_{1}, l_{2}, s}(\chi)\,,
\end{align}
with
\begin{align}
\tilde{\gamma}_{l_{1}, l_{2}, s}(\chi) &= \frac{2^{s - \frac{1}{6}(l_{1} - l_{2}) - 2}}{\pi} \Gamma\left(l_{2} - l_{1}\right) \Gamma(s) 
\nn \\
&\times \chi^{1 + \frac{1}{6}(l_{1} - l_{2}) - s} e^{{\cal{F}}(l_{1}, l_{2}, s; \chi)}\,.
\end{align}
The function ${\cal{F}}(l_{1}, l_{2}, s; \chi)$ is a Laurent series in $\chi$, specifically
\begin{equation}
{\cal{F}}(l_{1}, l_{2},s; \chi) = \sum_{k=0} {\cal{F}}_{k}(l_{1}, l_{2}, s) \chi^{-k}\,,
\end{equation}
where
\allowdisplaybreaks[4]
\begin{align}
{\cal{F}}_{0} &= \frac{i\pi}{12} \left(l_{1} + l_{2} - 6s\right)\,,
\\
{\cal{F}}_{1} &= \frac{i}{36} \left[l_{1}^{2} + 6 l_{2} - l_{2}^{2} + l_{1} (6-12s) + 36 s (s-1)\right]\,,
\\
{\cal{F}}_{2} &= \frac{1}{324} \left[l_{1}^{3} - 18 l_{2} + 9 l_{2}^{2} - l_{2}^{3} + l_{1}^{2}(9-18s) 
\right.
\nn \\
&\left.
- 108 s (1 - 3s + 2s^{2}) + 18 l_{1}(1-6s+6s^{2})\right]\,,
\end{align}
This completes the asymptotic expansion of the hypergeometric function.
\section{Asymptotic Expansion of Amplitudes}
\label{amps}

We here provide the asymptotic expansions of the amplitude functions $\tilde{{\cal{A}}}_{l_{1}, l_{2}, s}$. To simplify the expressions, we define $\bar{{\cal{A}}}_{l_{1}, l_{2}, s}$ such that
\begin{equation}
\tilde{{\cal{A}}}_{l_{1}, l_{2}, s} (\chi) = e^{i {\cal{E}}(\chi)} \bar{{\cal{A}}}_{l_{1}, l_{2}, s}(\chi)
\end{equation}
where
\begin{equation}
{\cal{E}}(\chi) = - \frac{\chi}{2} \left[-2 + \ln\left(\frac{9\chi^{2}}{4\zeta^{3}}\right)\right]
\end{equation}
and the non-zero $\bar{{\cal{A}}}$ functions are to ${\cal{O}}(\chi^{2})$
\allowdisplaybreaks[4]
\begin{widetext}
\begin{align}
\bar{\cal{A}}_{1,5,1/2}^{\times} &= \frac{(\frac{1}{108} + \frac{i}{108}) c_{2\beta} c_{\iota} (1 - e_{0}^{2})^{3/4} (-1 + 36 i \chi + 648 \chi^{2})}{\sqrt{6 \pi} \chi^{2} \zeta^{3/4}}\,,
\\
\bar{{\cal{A}}}_{10,8,1/2}^{\times} &= \frac{(\frac{1}{810} + \frac{i}{810}) c_{2\beta} c_{\iota} (1 - e_{0}^{2})^{5/4}}{\sqrt{6 \pi} \chi^{2} \chi_{\rm orb} \zeta^{15/4}} \left[(-63778 i + 24165 \chi + 19440 i \chi^{2} + 21870 \chi^{3}) \chi_{\rm orb}^{2} 
\right.
\nn \\
&\left.
+ 30 (-847 i + 342 \chi + 162 i \chi^{2}) \zeta^{3}\right]\,,
\\
\bar{{\cal{A}}}_{10,8,-1/2}^{\times} &= -\frac{(\frac{1}{108} + \frac{i}{108}) c_{2\beta} c_{\iota} (1 - e_{0}^{2})^{5/4} (-847 i + 342 \chi + 162 i \chi^{2}) (9 \chi_{\rm orb}^{2} + 4 \zeta^{3})}{\sqrt{6 \pi} \chi^{2} \chi_{\rm orb} \zeta^{15/4}}\,,
\\
\bar{{\cal{A}}}_{2,4,1/2}^{\times} &= \frac{(\frac{1}{54} - \frac{i}{54}) c_{\iota} (-2 + e_{0}^{2}) s_{2\beta} (-1 - 18 i \chi + 162 \chi^{2})}{(1 - e_{0}^{2})^{1/4} \sqrt{6 \pi} \chi^{2} \zeta^{3/4}}\,,
\\
\bar{{\cal{A}}}_{4,8,1/2}^{\times} &= \frac{(\frac{1}{2430} + \frac{i}{2430}) c_{\iota} (1 - e_{0}^{2})^{1/4}}{\sqrt{6 \pi} \chi^{2} \chi_{\rm orb} \zeta^{9/4}} \left\{(-1 + e_{0}^{2}) s_{2\beta} (-553 i + 1485 \chi + 12150 i \chi^{2} + 21870 \chi^{3}) \chi_{\rm orb} 
\right.
\nn \\
&\left.
+ 270 c_{2\beta} \sqrt{1 - e_{0}^{2}} \left[(-101 i + 90 \chi - 162 i \chi^{2}) \chi_{\rm orb}^{2} + 4 (-13 i + 9 \chi) \zeta^{3}\right]\right\}\,,
\\
\bar{{\cal{A}}}_{4,8,-1/2}^{\times} &= -\frac{(\frac{1}{9} + \frac{i}{9}) c_{2\beta} c_{\iota} (1 - e_{0}^{2})^{3/4} (-13 i + 9 \chi) (9 \chi_{\rm orb}^{2} + 4 \zeta^{3})}{\sqrt{6 \pi} \chi^{2} \chi_{\rm orb} \zeta^{9/4}}\,,
\\
\bar{{\cal{A}}}_{5,7,1/2}^{\times} &= \frac{(\frac{1}{19440} + \frac{i}{19440}) c_{\iota}}{(1 - e_{0}^{2})^{1/4} \sqrt{6 \pi} \chi^{2} \chi_{\rm orb} \zeta^{9/4}} \left\{c_{2\beta} (-1 + e_{0}^{2}) (3181 + 3510 i \chi + 68040 \chi^{2} - 174960 i \chi^{3}) \chi_{\rm orb} \sqrt{1 - e_{0}^{2}} 
\right.
\nn \\
&\left.
- 270 (-2 + e_{0}^{2}) s_{2\beta} \left[(563 + 252 i \chi + 648 \chi^{2}) \chi_{\rm orb}^{2} + 8 (29 + 18 i \chi) \zeta^{3}\right]\right\}\,,
\\
\bar{{\cal{A}}}_{5,7,-1/2}^{\times} &= \frac{(\frac{1}{36} + \frac{i}{36}) c_{\iota} (-2 + e_{0}^{2}) s_{2\beta} (29 + 18 i \chi) (9 \chi_{\rm orb}^{2} + 4 \zeta^{3})}{(1 - e_{0}^{2})^{1/4} \sqrt{6 \pi} \chi^{2} \chi_{\rm orb} \zeta^{9/4}}\,,
\\
\bar{{\cal{A}}}_{7,11,1/2}^{\times} &= -\frac{(\frac{1}{6480} + \frac{i}{6480}) c_{\iota} (1 - e_{0}^{2})^{5/4} s_{2\beta}}{\sqrt{6 \pi} \chi^{2} \chi_{\rm orb} \zeta^{15/4}} \left[(-356639 - 164970 i \chi + 184680 \chi^{2} - 174960 i \chi^{3}) \chi_{\rm orb}^{2} 
\right.
\nn \\
&\left.
+ 60 (-3169 - 1260 i \chi + 648 \chi^{2}) \zeta^{3}\right]\,,
\\
\bar{{\cal{A}}}_{7,11,-1/2}^{\times} &= \frac{(\frac{1}{432} + \frac{i}{432}) c_{\iota} (1 - e_{0}^{2})^{5/4} s_{2\beta} (-3169 - 1260 i \chi + 648 \chi^{2}) (9 \chi_{\rm orb}^{2} + 4 \zeta^{3})}{\sqrt{6 \pi} \chi^{2} \chi_{\rm orb} \zeta^{15/4}}\,,
\\
\bar{{\cal{A}}}_{1,5,1/2}^{+} &= \frac{(\frac{1}{216} + \frac{i}{216}) (1 + c_{\iota}^{2}) (1 - e_{0}^{2})^{3/4} s_{2\beta} (-1 + 36 i \chi + 648 \chi^{2})}{\sqrt{6 \pi} \chi^{2} \zeta^{3/4}}\,,
\\
\bar{{\cal{A}}}_{10,8,1/2}^{+} & = \frac{(\frac{1}{1620} + \frac{i}{1620}) (1 + c_{\iota}^{2}) (1 - e_{0}^{2})^{5/4} s_{2\beta}}{\sqrt{6 \pi} \chi^{2} \chi_{\rm orb} \zeta^{15/4}} \left[(-63778 i + 24165 \chi + 19440 i \chi^{2} + 21870 \chi^{3}) \chi_{\rm orb}^{2} 
\right.
\nn \\
&\left.
+ 30 (-847 i + 342 \chi + 162 i \chi^{2}) \zeta^{3}\right]\,,
\\
\bar{{\cal{A}}}_{10,8,-1/2}^{+} &= -\frac{(\frac{1}{216} + \frac{i}{216}) (1 + c_{\iota}^{2}) (1 - e_{0}^{2})^{5/4} s_{2\beta} (-847 i + 342 \chi + 162 i \chi^{2}) (9 \chi_{\rm orb}^{2} + 4 \zeta^{3})}{\sqrt{6 \pi} \chi^{2} \chi_{\rm orb} \zeta^{15/4}}\,,
\\
\bar{{\cal{A}}}_{2,4,1/2}^{+} &= \frac{(\frac{1}{108} + \frac{i}{108}) \left[c_{2\beta} (1 + c_{\iota}^{2}) (-2 + e_{0}^{2}) + e_{0}^{2} s_{\iota}^{2}\right] (-i + 18 \chi + 162 i \chi^{2})}{(1 - e_{0}^{2})^{1/4} \sqrt{6 \pi} \chi^{2} \zeta^{3/4}}\,,
\\
\bar{{\cal{A}}}_{4,8,1/2}^{+} &= \frac{(\frac{1}{4860} + \frac{i}{4860}) (1 + c_{\iota}^{2}) (1 - e_{0}^{2})^{1/4}}{\sqrt{6 \pi} \chi^{2} \chi_{\rm orb} \zeta^{9/4}} \left\{-c_{2\beta} (-1 + e_{0}^{2}) (-553 i + 1485 \chi + 12150 i \chi^{2} + 21870 \chi^{3}) \chi_{\rm orb} 
\right.
\nn \\
&\left.
+ 270 \sqrt{1 - e_{0}^{2}} s_{2\beta} \left[(-101 i + 90 \chi - 162 i \chi^{2}) \chi_{\rm orb}^{2} + 4 (-13 i + 9 \chi) \zeta^{3}\right]\right\}\,,
\\
\bar{{\cal{A}}}_{4,8,-1/2}^{+} &= -\frac{(\frac{1}{18} + \frac{i}{18}) (1 + c_{\iota}^{2}) (1 - e_{0}^{2})^{3/4} s_{2\beta} (-13 i + 9 \chi) (9 \chi_{\rm orb}^{2} + 4 \zeta^{3})}{\sqrt{6 \pi} \chi^{2} \chi_{\rm orb} \zeta^{9/4}}\,,
\\
\bar{{\cal{A}}}_{5,7,1/2}^{+} &= \frac{(\frac{1}{38880} + \frac{i}{38880})}{(1 - e_{0}^{2})^{1/4} \sqrt{6 \pi} \chi^{2} \chi_{\rm orb} \zeta^{9/4}} \left\{-(1 + c_{\iota}^{2}) (1 - e_{0}^{2})^{3/2} s_{2\beta} (3181 + 3510 i \chi + 68040 \chi^{2} - 174960 i \chi^{3}) \chi_{\rm orb} 
\right.
\nn \\
&\left.
+ 270 \left[c_{2\beta} (1 + c_{\iota}^{2}) (-2 + e_{0}^{2}) + e_{0}^{2} s_{\iota}^{2}\right] \left[(563 + 252 i \chi + 648 \chi^{2}) \chi_{\rm orb}^{2} + 8 (29 + 18 i \chi) \zeta^{3}\right]\right\}\,,
\\
\bar{{\cal{A}}}_{5,7,-1/2}^{+} &= \frac{(\frac{1}{72} - \frac{i}{72}) \left[c_{2\beta} (1 + c_{\iota}^{2}) (-2 + e_{0}^{2}) + e_{0}^{2} s_{\iota}^{2}\right] (-29 i + 18 \chi) (9 \chi_{
\rm orb}^{2} + 4 \zeta^{3})}{(1 - e_{0}^{2})^{1/4} \sqrt{6 \pi} \chi^{2} \chi_{\rm orb} \zeta^{9/4}}\,,
\\
\bar{{\cal{A}}}_{7,11,1/2}^{+} &= -\frac{(\frac{1}{12960} + \frac{i}{12960}) c_{2\beta} (1 + c_{\iota}^{2}) (1 - e_{0}^{2})^{5/4}}{\sqrt{6 \pi} \chi^{2} \chi_{\rm orb} \zeta^{15/4}} \left[(356639 + 164970 i \chi - 184680 \chi^{2} + 174960 i \chi^{3}) \chi_{\rm orb}^{2} 
\right.
\nn \\
&\left.
+ 60 (3169 + 1260 i \chi - 648 \chi^{2}) \zeta^{3}\right]\,,
\\
\bar{{\cal{A}}}_{7,11,-1/2}^{+} &= -\frac{(\frac{1}{864} + \frac{i}{864}) c_{2\beta} (1 + c_{\iota}^{2}) (1 - e_{0}^{2})^{5/4} (-3169 - 1260 i \chi + 648 \chi^{2}) (9 \chi_{\rm orb}^{2} + 4 \zeta^{3})}{\sqrt{6 \pi} \chi^{2} \chi_{\rm orb} \zeta^{15/4}}\,.
\end{align}
\end{widetext}
%
\bibliography{master}
\end{document}